\documentclass[epj]{svjour}
\usepackage{latexsym}
\usepackage{graphics}
\usepackage{lmodern}

\usepackage[T1]{fontenc}
\usepackage[utf8]{inputenc}
\usepackage[english]{babel}
\usepackage{geometry}
\usepackage{cprotect}
\usepackage{float}
\usepackage{amsmath}
\usepackage{amssymb}
\usepackage{graphicx}
\usepackage{microtype}
\usepackage[unicode=true,pdfusetitle,
 bookmarks=true,bookmarksnumbered=false,bookmarksopen=false,
 breaklinks=false,pdfborder={0 0 1},backref=false,colorlinks=false]
 {hyperref}
\usepackage{colortbl}
\begin{document}
\title{A colloquium on the variational method applied to excitons in 2D materials}
\author{M. F. C. Martins Quintela\inst{1}, N. M. R. Peres\inst{1,2}}
\authorrunning{M. M. Quintela, N. M. R. Peres}
\titlerunning{}
\offprints{}          % Insert a name or remove this line
\institute{Centro de F\'{i}sica and Departamento de F\'{i}sica and
QuantaLab, Universidade do Minho, P-4710-057 Braga, Portugal \and International Iberian Nanotechnology Laboratory
(INL), Av Mestre Jos\'{e} Veiga, 4715-330 Braga, Portugal}
\date{Received: date / Revised version: date}
\abstract{
In this colloquium, we review the research on excitons in van der Waals
heterostructures from the point of view of variational calculations. We first make a presentation of the current and past literature,
followed by a discussion on the connections between experimental and theoretical results. In particular, we focus our review of the literature on the absorption spectrum and polarizability, as well as the
Stark shift and the dissociation rate. Afterwards, we begin the discussion of the
use of variational methods in the study of excitons. We initially
model the electron-hole interaction as a soft-Coulomb potential,
which can be used to describe interlayer excitons. Using an \emph{ansatz}, based on the solution
for the two-dimensional quantum harmonic oscillator, we study the
Rytova-Keldysh potential, which is appropriate to describe intralayer excitons in two-dimensional (2D) materials. These variational energies are then recalculated with a different \emph{ansatz}, based on the exact wavefunction of the 2D hydrogen atom, and the obtained energy curves are compared. Afterwards, we discuss the
Wannier-Mott exciton model, reviewing it briefly before focusing on
an application of this model to obtain both the exciton absorption
spectrum and the binding energies for certain values of the physical 
parameters of the materials. Finally, we briefly discuss an
approximation of the electron-hole interaction in interlayer excitons as an harmonic potential
and the comparison of the obtained results with the existing values
from both first--principles calculations and experimental measurements.
%
%\PACS{
%      {PACS-key}{discribing text of that key}   \and
%      {PACS-key}{discribing text of that key}
%     } % end of PACS codes
} %end of abstract
\global\long\def\ket#1{\left|#1\right\rangle }%
\global\long\def\bra#1{\left\langle #1\right|}%
\global\long\def\braket#1#2{\left\langle #1|#2\right\rangle }%
\maketitle
\section{Introduction}
\label{intro}
Alongside graphene, a wide range of bi-dimensional materials are currently
studied and have a plethora of different physical properties and applications\cite{Mounet_2018}.
An important subclass of these materials is the family of transition-metal
dichalcogenides with chemical formula $\mathrm{MX_{2}}$, where $\mathrm{M}$
is a transition metal and $\mathrm{X}$ a chalcogen atom. These materials,
specifically those with group-VI transition metals, are semiconductors
which exhibit strong light-matter coupling, as well as having direct
band gaps in the infrared and visible spectral regimes. Having been
studied in their bulk form since the 1960's\cite{1963,Fortin_1982},
the advent of the study of two-dimensional (2D) layers with atomic scale thickness renewed
the interest in these materials and their properties that make them good candidates
for various applications in optics and optoelectronics\cite{mueller_exciton_2018}.

\begin{figure}[H]
\centering
\includegraphics[scale=0.3]{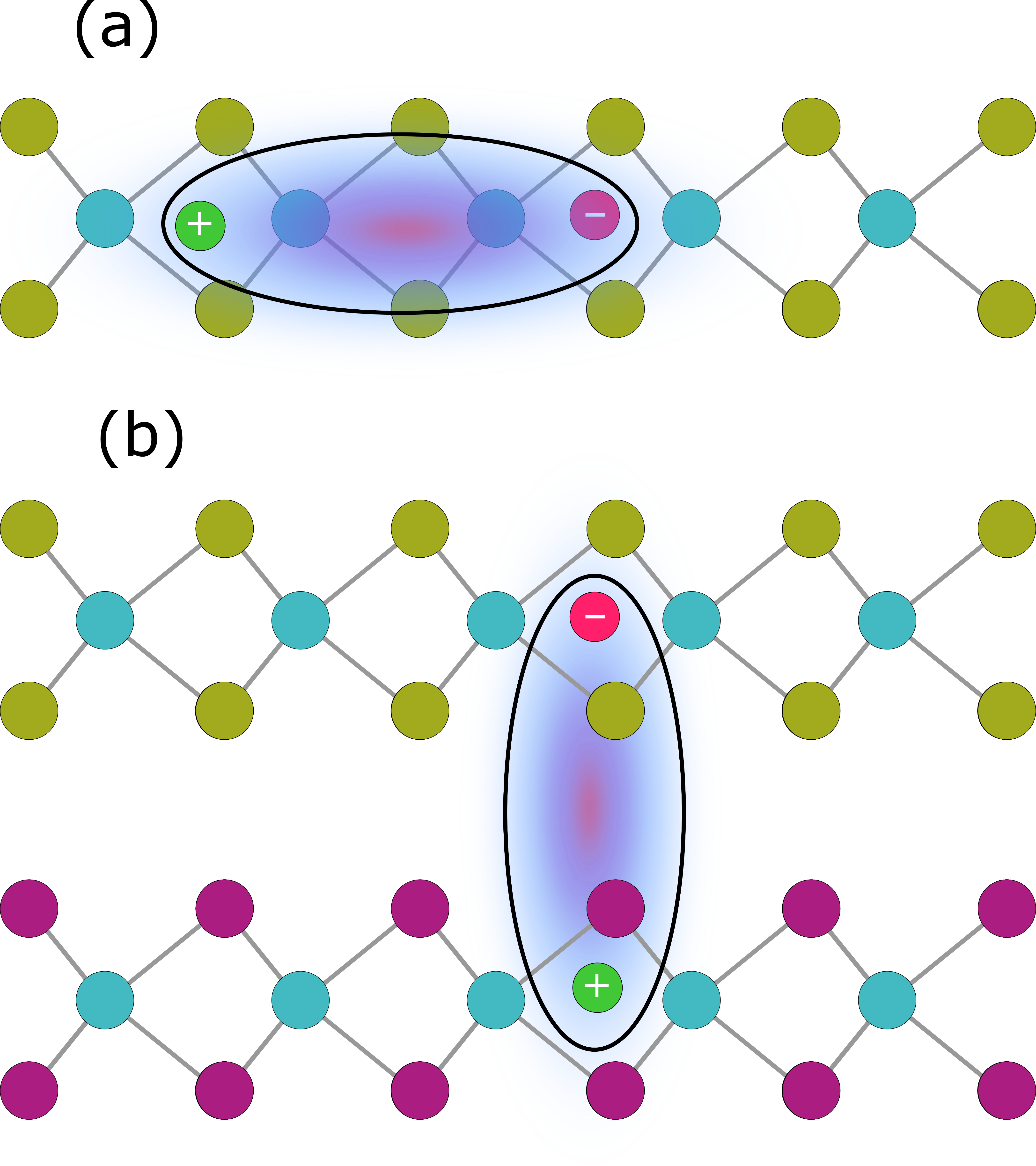}
\caption{{(a) Intralayer excitons in a TMD; (b) Interlayer excitons in two different TMDs.}}\label{fig:direct_indirect_exciton}
\end{figure}

The optical response of these semiconductors is mainly dominated by
the excitation of electrons from the valence band to the conduction
band\cite{RevModPhys.90.021001,nwu078}. Such a phenomena can be described by a pair of interacting (effective)
particles, one being a conduction electron and the other being a hole
left in the valence band, with opposite charge to the electron (Figure \ref{fig:direct_indirect_exciton}--a). Considering
frequencies above the band gap, a transition from the valence to the
conduction band is possible, which means that the absorption becomes
finite. For certain materials, absorption peaks can be measured below
the band gap, which can be explained by the presence of excitonic
states (bound states of the said electron and hole).

As the electron and hole are of opposite charges, the most natural
formulation of their interaction will be an attractive interaction. This
will lead to the possibility of the formation of bound states between
these particles, analogous to those formed between an electron and
a proton in the hydrogen atom\cite{Knox_1983}. Unfortunately, the
small effective mass of the particles in question and large screening effects
means that the excitonic binding energy in bulk materials is of the order of 
$\mathrm{meV}$, while the room-temperature thermal fluctuations
are about $25\,\mathrm{meV}$. These fluctuations mask excitonic effects
unless the material is sufficiently cooled down.

The hydrogen atom and, consequently, hydrogen-like problems are some
of the most studied systems in physics. While the simplest model consisting
only of the Coulomb interaction has a well known exact solution, when
one wishes to introduce more complex interactions no exact solutions
are known. This becomes increasing problematic when one wishes to
study systems with local/non-local screening, or finite (and externally
fixed) minimum separation between particles.

In 2D transition-metal dichalcogenides (TMDs), where the screening effect
is reduced relatively to their bulk counterparts, the exciton binding energies reach values the order of
$100\,\mathrm{meV}$. Therefore, these class of excitons are more
easily accessible to experimental study, as they are observable at
room temperature. As an example, $\mathrm{WSe}_{2}$ in a fused quartz
substrate presents two excitonic peaks in the linear absorption at
$1.65$ and $2.08\,\mathrm{eV}$. The presence of two peaks instead of a single one is the signature  of strong spin-orbit coupling in this system. The position of these peaks in frequency, however, was not consistent
with the bare Coulomb interaction, which shows the necessity of including
screening effects in the electron-hole interaction.

The simplest way of including a screening-like effect is via the soft-Coulomb
potential, which introduces a material-dependent minimum separation-like
parameter between the electron and the hole. Although this approach might seem somewhat artificial, it is relevant to the discussion of interlayer excitons (see below). Another of these potentials
is the Rytova-Keldysh potential\cite{rytova1967,1979JETPL..29..658K},
which includes a material dependent parameter and reduces to the Coulomb
potential in the regime when this parameter is zero. Mathematically,
this potential is significantly more complex than the Coulomb potential,
which in turn further complicates the analytical work in determining
the absorption spectrum. It can, however, be closely approximated
by simpler functions, which might help with the calculations for the
absorption spectrum.

Another relevant system is that of spatially separated excitons (also called
indirect excitons, although this terminology may be misleading), where each element of the electron-hole pair is
situated in a different layer, creating a minimum distance between
the electron and the hole (Figure \ref{fig:direct_indirect_exciton}--b). These excitons are specially relevant in
van der Waals (vdW) heterostructures, where multiple layers of different
TMDs are stacked\cite{Geim_2013,Dong_2019}. With the previously-mentioned soft-Coulomb potential,
this bias distance is easily introduced as a minimum separation parameter,
which depends on the specific heterostructure. The obtained theoretical
results (from either variational or numerical methods) can be compared
with experimental results, where the minimum separation is usually
of the order of the effective Bohr radius for an electron-hole pair
in the material in question.

Besides vdW heterostructures, where the screening effects
reflect the nature of the different materials in each layer, these
same screening effects are fundamental when one wishes to study TMD
monolayers surrounded by a dielectric medium with a dielectric constant
different from that of the TMD\cite{Thygesen_2017}. In these systems, as mentioned before,
the physics of the electron-hole is better modeled by replacing the
Coulomb potential with the Rytova-Keldysh potential \cite{rytova1967,1979JETPL..29..658K}
with parameters related to the relative dielectric constants of the
media surrounding the monolayer, as well as the 2D polarizability
of the TMD \cite{brunetti_optical_2018,van_tuan_coulomb_2018,zhang_two-dimensional_2019,Scharf_2019,kamban_interlayer_2020,Cavalcante_2018}.

\section{Brief Historical Review}
\label{sec:1}
After the first mechanical exfoliation of graphene in 2004 by Novoselov
and Geim\cite{novoselov_electric_2004}, the field of two-dimensional
materials boomed. Just one year after the exfoliation of graphene, in 2005, Novoselov \emph{et al.}\cite{novoselov_two-dimensional_2005} showed that
the same technique could be applied to isolate other monolayer materials.
This was specifically done for a 2D semiconductor, molybdenum disulfide
($\mathrm{MoS_{2}}$). Despite these advances, most of the research
on 2D materials was focused on the unknown properties of graphene.
This changed somewhat in 2010, when Heinz and Wang demonstrated \cite{Mak2010,wang_2010}
(independently) that a $\mathrm{MoS_{2}}$ monolayer exhibits strong
photoluminescence emission and has a direct band gap. The bandgaps in the near-infrared to visible range, the high photo-catalytic and mechanic stability, the decent charge carrier mobility, and the presence of exotic many-body phenomena make these materials extremely interesting from a fundamental research, device application, and innovation perspectives\cite{Wurstbauer_2017}. Despite the
quick recognition that the optical response of transition metal dichalcogenides arises from excitons, quantitative measurements of the binding
energy were only reported around 2014 \cite{chernikov_2014}. Excitons
in these materials are tightly bound, and dominate the optical response
even at room-temperature \cite{Mak_2012,He_2014}. This possibility
comes about due to the reduced screening among the electrons in the
TMDs, which has its origin in the low dimensionality of the 2D materials.

The demonstration of valley-selective optical excitation, in 2012,
brought forth the field of TMD-based valleytronics \cite{Xiao_2012}.
In 2013 \cite{Geim2013}, the advent of vdW heterostructures
allowed the fabrication of TMD heterostructures, and consequently
the observation of interlayer excitons. The more recent developments
of optoelectronic devices, whose structure resembles that of the first
$\mathrm{MoS_{2}}$ monolayer field-effect transistor (demonstrated
in 2011 by Kis \emph{et al}\cite{kis_2011}) has happened in parallel
with the research on the optical properties of TMDs. This same structure
was used to realized atomically thin phototransistors one year later
\cite{Yin_2011}.

The realization of atomically thin \emph{p-n} junctions for both vertical
\cite{Pospischil_2014} and lateral \cite{Furchi2014} geometries,
in 2014, lead to the realization of light-emitting diodes, solar cells,
and photodiodes. The potential efficiency of these devices was greatly
improved by the 2015 demonstration of near-unity photon quantum yield
in a TMD monolayer \cite{Amani_2015}. More recently, in 2017, the
role of dark exciton states has become a major field of research,
with optically allowed and forbidden dark excitons forming due to
weak dielectric screening and strong geometrical confinement in TMD-based
vdW heterostructures \cite{Zhang_2017,Molas_2017}.

Recently, the calculation and the experimental measurements of the
Stark-shift and ionization rate of excitons in TMD has attracted considerable
attention \cite{scharf_excitonic_2016,haastrup_stark_2016,massicotte_dissociation_2018,Rigosi_2015}.
The possibility of increasing the efficiency of photodetectors using
the Stark effect has grown the interest in this topic. However, the
same class of problems is largely understudied in vdW heterostructures,
with only one paper published so far on this topic \cite{kamban_interlayer_2020}.

The interest in the optical and excitonic properties of atomically
thin TMDs \cite{Berkelbach_2018} has peaked in recent years, with
numerous works focused on the study of both the excitonic binding
energies \cite{aquino_accurate_2005,scharf_excitonic_2016,grasselli_variational_2017,inarrea_effects_2019},
exciton-exciton/exciton-electron interactions \cite{Shahnazaryan_2017},
strain effects \cite{Aas_2018}, their presence in vdW twisted heterostructures \cite{Jin_2019}, anisotropic semiconductors \cite{Castellanos_Gomez_2015} and quantum dots \cite{Li_2015,Pawbake_2016,Choi_2017,Ratnikov_2020}, and several magneto-optical effects
\cite{chaves_excitonic_2017,have_excitonic_2019,henriques_excitonic_2020,henriques_optical_2020}
published by many. 

In what follows, we will briefly review some of the various approaches used in the
study of excitons in vdW heterostructures, starting from the study of the electron-hole interaction
in both monolayer TMDs \cite{van_tuan_coulomb_2018,durnev_excitons_nodate}
and vdW heterostructures \cite{van_der_donck_excitons_2018,brunetti_optical_2018,Castellanos_Gomez_2014}, as well as the study of screening effects in the interaction among the charge carriers \cite{Zhang_2019,Scharf_2019}.

\section{Connection with Experimental Results}
\label{sec:2}
Discovered in 1913 by Stark\cite{Stark_1914}, the Stark effect consists
in the splitting of the spectral lines due to the presence of an external
static electric field (analogously to the Zeeman effect). This effect is
greatly enhanced by excitons in vdW heterostructures, as the electron
and the hole that form the exciton are ``pulled'' in opposing directions
by the external field, but remain confined in the material until the field is strong enough to dissociate it.

For weak electric fields, the Stark shifts of electrons vary, in accordance
with perturbation theory, approximately quadratically with the external
field $\mathcal{E}$, i.e., $E\approx E_{0}-\frac{1}{2}\alpha\mathcal{E}^{2}$,
where $E_{0}$ is the unperturbed energy and $\alpha$ the in-plane
polarizability. This allows for the calculation of the polarizability,
which can then be used as a comparison against experimental results.
The calculation of the polarizabilities reveals that they are significantly
larger (around $3\times$ greater) for interlayer excitons, compared
to their monolayer counterparts \cite{kamban_interlayer_2020}. This
phenomena in bilayer vdW heterostructures can be explained
by the increased screening and the vertical separation of the electron-hole
pair, both of which reduce the binding energy and, therefore, facilitate
the polarization of the exciton.

As the electron and the hole are pulled in opposing directions by
the external field, the exciton may dissociate. This dissociation
is realized by a non-vanishing imaginary part of the energy eigenvalue.
Knowing the imaginary part of the energy eigenvalue, one can quickly
calculate the field dissociation rate as $\Gamma=-2\,\mathrm{Im}E/\hbar$.
Kamban and Pedersen\cite{kamban_interlayer_2020} obtain this imaginary
part by applying the external complex scaling method \cite{Simon_1979,McCurdy_1991},
i.e., by rotating the radial coordinate into the complex plane by
an angle $\phi$ outside of a specific radius $R$. The equation for
the eigenstate is then split into the radial and the angular parts.
The former is dealt with using a finite element basis consisting of
Legendre polynomials, while the latter is solved using a cosine basis.

\begin{figure}
\centering
\includegraphics[scale=0.3]{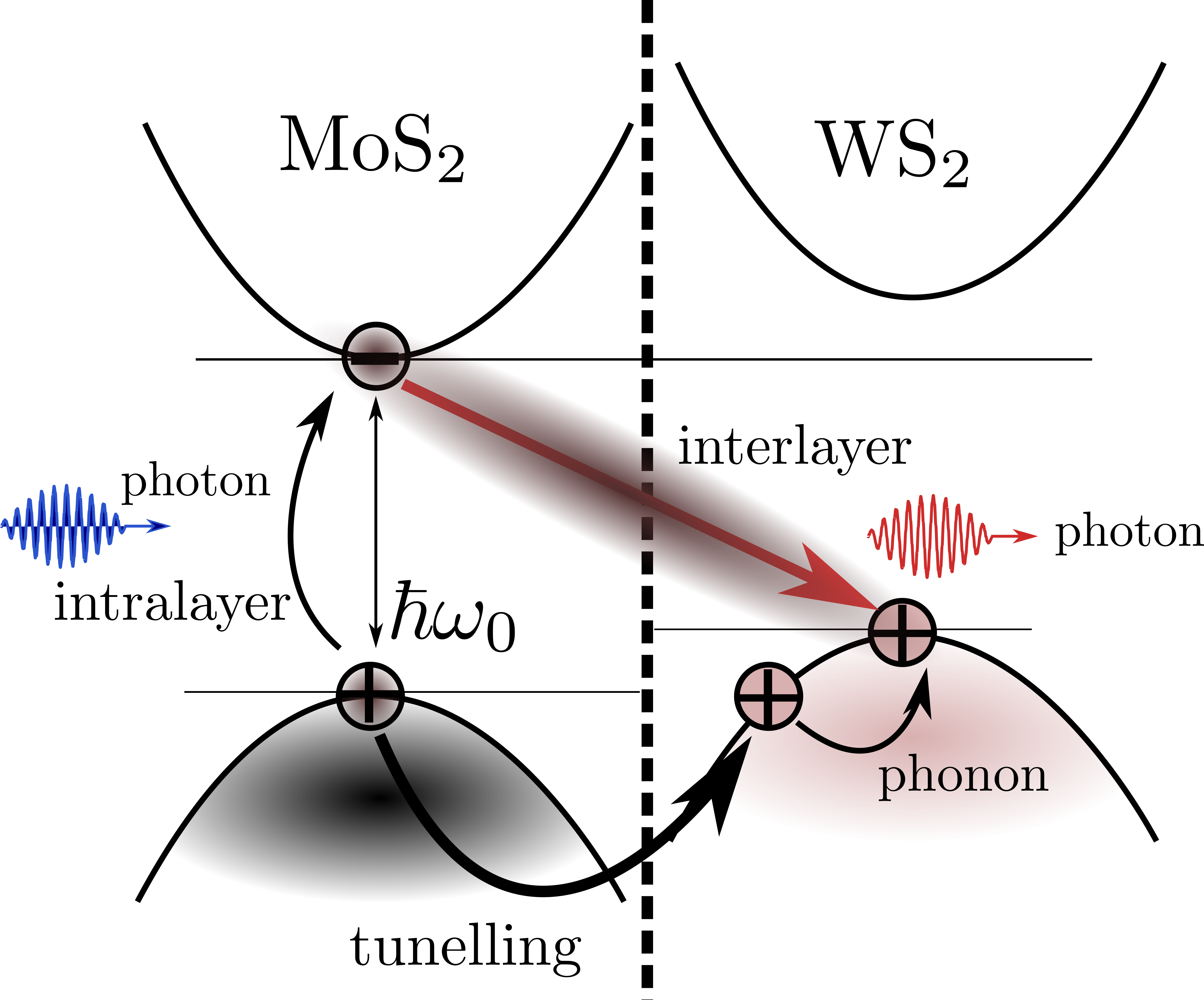}
\caption{Mechanism for interlayer exciton formation: (i) A photon of energy $\hbar\omega_0$ is absorbed in MoSe$_2$ creating a intralayer exciton; (ii) the hole thus tunnels to the valence band of the nearby WSe$_2$; (iii) due to phonon absorption the hole moves to the top of the valence band of  WSe$_2$; (iv) the strong electrostatic interaction between the electron and hole in the two different layers forms an intralayer exciton.
(This image is adapted from Ref. \cite{Malic-2019}.)
}\label{fig:interlayer_exciton}
\end{figure}

The dissociation times
can be approximated as $\tau=1/\Gamma$ and are, at least, $10$ orders
of magnitude greater than the experimentally obtained $50\,\mathrm{fs}$ required for holes to tunnel into the $\mathrm{WS_{2}}$ layer of a $\mathrm{MoS_{2}}/\mathrm{WS_{2}}$ heterostructure \cite{Hong_2014}. This time
will likely be affected by material-- (and medium) specific parameters, as these influence both the band gap and the binding energies of the excitons \cite{Meckbach_2018,Kumar_2018}. 
This large discrepancy is explained
by the transition from intralayer to interlayer excitons (due to the staggered/type-II band alignment, see Fig. \ref{fig:interlayer_exciton}) before the
field can dissociate these same excitons. Also, the luminescence exhibited by these interlayer excitons is exhibited at energies lower than that for intralayer complexes \cite{Vialla_2019}, as is clear from the band diagram of Fig. \ref{fig:interlayer_exciton}.

After tunneling, interlayer excitons have long lifetime due to the small overlap of the individual electron and hole wave functions \cite{Miller_2017}. As such, when an in-plane field is applied, intralayer excitons will tunnel into interlayer excitons with sufficiently long lifetimes for them to dissociate. The much larger dissociation rates of interlayer excitons when compared to intralayer ones corroborates this, with $\Gamma\sim10^{4}\,\mathrm{s}^{-1}$ for interlayer excitons in a freely suspended (i.e., not surrounded by a dielectric medium) $\mathrm{MoS_{2}}/\mathrm{WS_{2}}$ heterostructure, but only $\Gamma\sim10^{-38}\,\mathrm{s}^{-1}$
for monolayer $\mathrm{MoS_{2}}$ (at $\mathcal{E}=10\,\mathrm{V}/\mu \mathrm{m}$)\cite{Kamban_2019}. 

Furthermore, for the same electric field intensity, the
dissociation rates for intralayer excitons in the top and bottom layers
of $\mathrm{MoS_{2}}/\mathrm{WS_{2}}$ structures differ by seven
orders of magnitude ($\Gamma\sim2.1\times10^{-3}\,\mathrm{s}^{-1}$ and $\Gamma\sim2.9\times10^{4}\,\mathrm{s}^{-1}$,
respectively), which can be explained from the reduced masses in each
material $\mu\approx0.25,\,0.15$ (in units of the bare electron mass), respectively.

When studying exciton dissociation it is also important to take into
account their radiative lifetimes. Wang \emph{et al.}\cite{Wang_2016}
focus their efforts on $\mathrm{Mo}\mathrm{S}_{2}$, calculating a
radiative lifetime of around $180-300\,\mathrm{fs}$, in good agreement
with recent experimental results for excitons with near-zero momentum
in $\mathrm{WSe_{2}}$, in the range $150-250\,\mathrm{fs}$ \cite{Moody_2015}.
Furthermore, Wang \emph{et al.}\cite{Wang_2016} show that excitons
with very long $\left(\geq1\,\mathrm{ns}\right)$ exciton lifetimes
that have been observed in clean $\mathrm{MoS_{2}}$ monolayers at
small photoexcited exciton densities \cite{Amani_2015} are consistent
with strongly localized excitons. 

The absorption spectrum is one of the most used physical quantity
of comparison between theoretical calculations and experimental results regarding both excitons \cite{Han_2018} and trions \cite{Courtade_2017}.
These theoretical calculations come from both analytical, quasi-analytical
and numerical approaches, usually expressed in terms of the optical
conductivity. Looking specifically at some recent works
regarding the calculation of the absorption spectrum, we mention Zhang
\emph{et al.}\cite{zhang_absorption_2014}, Van der Donck and Peeters\cite{van_der_donck_excitons_2017,van_der_donck_interlayer_2018},
and Henriques \emph{et al.}\cite{henriques_optical_2020}. Recent numerical
studies \cite{van_der_donck_excitons_2017} also focus on the importance
of considering a multiband model (including spin-orbit coupling) when
calculating the exciton/trion binding energies and their absorption
spectrum (a trion is a charged complex, which in a $n$-doped semiconductor is composed of the two electrons and a hole).

\begin{figure}[H]
\centering
\includegraphics[scale=0.25]{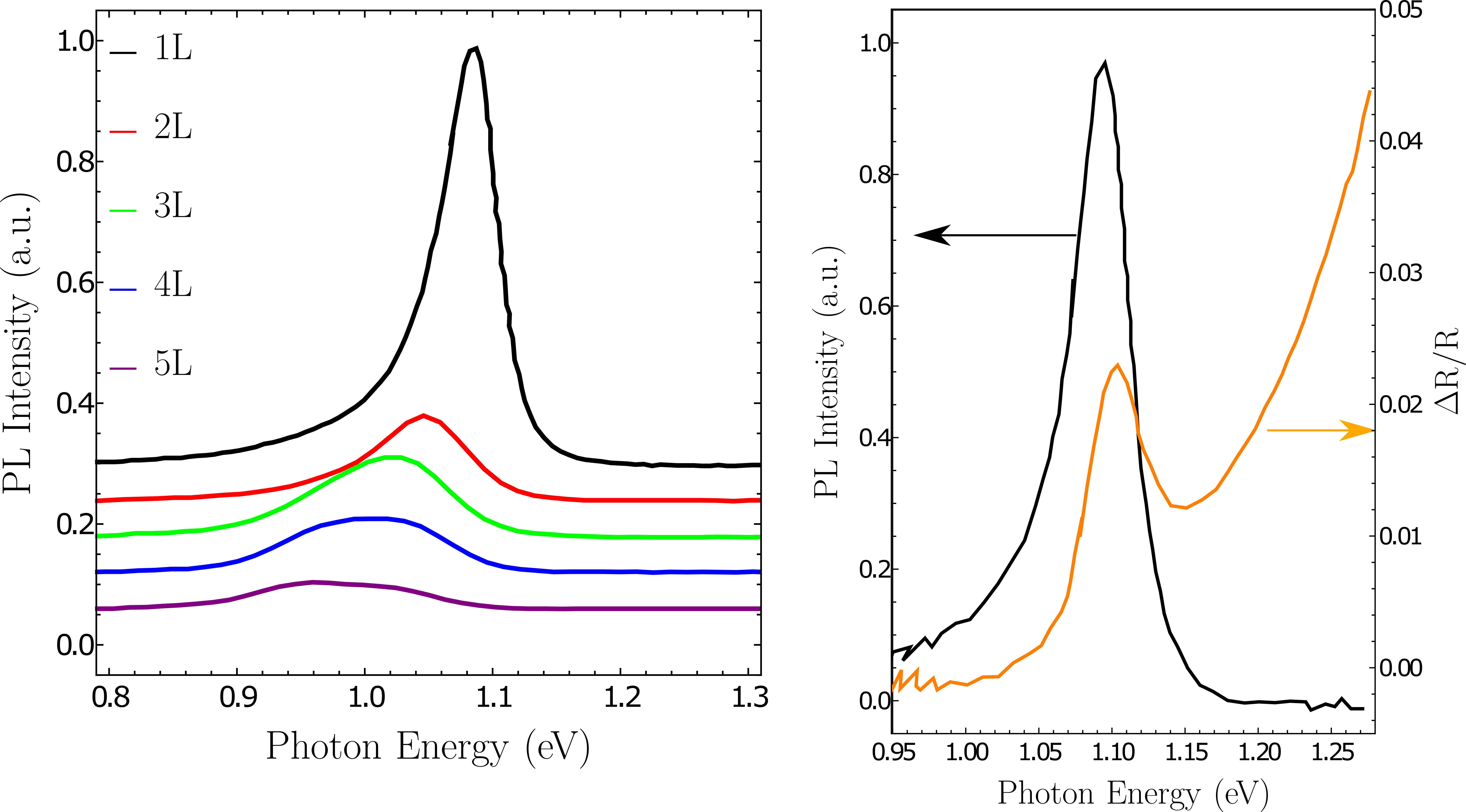}
\caption{Photoluminescence (PL) and absorbance spectra. (Left) PL spectrum 
for MoTe$_2$ multilayers. (Right) PL spectrum (black) and absorbance (orange)
for MoTe$_2$ single layer. 
The strong PL signal  for MoTe$_2$ single layer is evident.
(Figure adapted from Ref. \cite{Ruppert-2014}.)
}
\label{fig:PL}
\end{figure}

There are two complementary methods of accessing the optical properties of 2D materials: photoluminescence (PL) and absorption measurements ($\Delta R/R$). They both convey similar information. Ruppert  \emph{et al.}   \cite{Ruppert-2014} measured the PL and the absorption in MoTe$_2$ multilayers down to the single layer case. They showed that the PL signal increases significantly when one moves from the bulk crystal to the single layer limit (see Fig. \ref{fig:PL}). This led these authors to conclude that MoTe$_2$ single layer is a direct band gap material with a band-gap of 1.1 eV, thus extending the class of 2D direct band gap materials from the visible to the near-infrared.

\section{Variational Methods}

One of the simplest approach to study systems for which there is
no analytical solution is using variational methods, which allows one to
obtain approximations to both the wavefunction and its corresponding energy eigenvalue.
The obtained closed-form analytical solutions give greater physical
intuition on the result than numerical solutions. From the numerous
works by multiple authors that employ variational methods, we specifically
mention those by Grasselli\cite{grasselli_variational_2017}, Zhang
\emph{et al.}\cite{zhang_absorption_2014,zhang_two-dimensional_2019},
Semina\cite{Semina_2019}, Planelles\cite{planelles_simple_2017},
and Lundt \emph{et al.}\cite{Lundt_2018}.

This method is usually performed in one of three ways: by choosing
a trial wavefunction (\emph{ansatz}) depending on a series of parameters
and minimizing the expectation value of the Hamiltonian regarding
these same parameters, quantum Monte Carlo methods \cite{Vialla_2019}, or by imposing a set of boundary conditions
and constructing the wavefunction as a series expansion which obeys
these same conditions by definition (frequently used when dealing
with polygonal enclosures \cite{bhat_flexural_1987,liew_set_1991,Quintela_2020}).
This third procedure is not as usefull when one is dealing with infinite/semi--infinite
regions, and will therefore remain undiscussed.

Over the next two sections, we will briefly describe two distinct
approaches to variational methods, both based around choosing a trial
function to solve an eigenvalue problem where the exact solution is
either unknown or non--existent. Firstly, we will analyze
the process described by Grasselli\cite{grasselli_variational_2017},
applying it afterwards to both a different potential and a different
\emph{ansatz. }Secondly, we discuss a work by Zhang \emph{et al.}\cite{zhang_absorption_2014},
quickly reviewing the excitonic states, as well as the optical conductivity
and the absorption spectum, and finally comparing the exciton radii
and binding energies to those from first--principles approaches available
in the literature (see Sec. \ref{sec:binding_energy}).

\section{The Variational Approach to 2D systems}

The use of variational wave functions for describing excitonic properties has a long story in condensed matter and has been used in many systems, including quantum-well wires \cite{Brown_1987}. These same variational approaches have also been considered when it comes to the description of the excitonic Stark effect in confined systems \cite{Feng_1993}. These methods have also been applied to the study of biexcitons, quasi-particles consisting of two electrons and two holes \cite{Akimoto_1972}.

A simple example of variational methods is employed by Grasselli\cite{grasselli_variational_2017}
to obtain the variational energies of the soft-Coulomb
potential. Unlike the hydrogenic problem with the regular Coulomb
interaction, exact solutions of the Schr{\"o}dinger equation are not known for the soft-Coulomb potential. This potential is given by 
\begin{equation}
V_{\mathrm{s-C}}\left(r\right)=-\frac{k}{\sqrt{r^{2}+d^{2}}},
\end{equation}
where $k=e^{2}/\left(4\pi\epsilon_{0}\epsilon_{r}\right)$, $e$ is the electron charge, $\epsilon_0$ is the vacuum permittivity, and $\epsilon_{r}$
is the relative dielectric permittivity of the material, and is obtained
for the bare Coulomb potential by introducing a fixed bias distance
in the form of the parameter $d$. This potential is frequently used
in semiconductor physics, as it:
\begin{enumerate}
\item overcomes the divergence issues of the Coulomb potential at the origin
$\left(r=0\right)$, as well as the infinite binding energy of the
ground-state in one-dimension \cite{Loudon_1959}. In this case, $d\neq0$
is taken as a fixed cut-off parameter of the order of the screening length;
\item represents the coupling of an electron confined in a layer to a hole
sitting in a different layer, a distance $d$ apart.
\end{enumerate}
The method introduced by Grasselli will be presented in the context of a simple 2D material in Sec.
\ref{subsec:Soft-Coulomb-Results}, following which we will apply
the same methodology for the more difficult Rytova-Keldysh potential
in Sec. \ref{subsec:Rytova-Keldysh-Potential}.

\subsection{Interlayer Excitons in $\mathrm{hBN}$: the Soft--Coulomb Potential}\label{subsec:Soft-Coulomb-Results}

Few layers hexagonal Boron Nitride (hBN) hosts excitons in the ultraviolet. If the number of layers is small we can treat this system as coupled 2D layers. In this case the parameter $d$ is the distance between the layers where the electron and the hole are respectively located (interlayer excitons). The problem of excitons in a single layer of hBN was treated by Henriques et al \cite{Henriques_2019}.

Considering a two-dimensional system, the \emph{ans{\"a}tze} for
the ground- and first-excited states are based on the solutions for
the quantum harmonic oscillator and are given by\cite{grasselli_variational_2017}
\begin{align}
\phi_{0}\left(r\right) & =\sqrt{\frac{\beta_{0}}{\pi}}e^{-\beta_{0}r^{2}/2}, \\ \phi_{1x}\left(r,\theta\right) & =\sqrt{\frac{\beta_{1}}{\pi}}\sqrt{2\beta_{1}}r\cos\theta e^{-\beta_{1}r^{2}/2},
\end{align}
where $\beta_{0}$ and $\beta_{1}$ are variational parameters. The $y$--component for the first-excited state is, by symmetry,
degenerate and mutually orthogonal with the $x$--component and,
as such, will not be discussed.

The expectation values of the Hamiltonian
\begin{equation}
H\left(r\right) = - \frac{\hbar ^{2}}{2m}\nabla ^{2} + V\left(r\right)
\end{equation} 
taken with these two variational wave functions are given by 
\begin{align}
\varepsilon_{0}\left(\beta_{0},d\right) & =\frac{\hbar^{2}\beta_{0}}{2m}-k\sqrt{\beta_{0}\pi}e^{d^{2}\beta_{0}}\text{erfc}\left(d\sqrt{\beta_{0}}\right),\label{eq:grasselli_sc}\\
\varepsilon_{1}\left(\beta_{1},d\right) & =\frac{\hbar^{2}\beta_{1}}{m}-\frac{k}{2}\sqrt{\beta_{1}\pi}\left[2\frac{d\sqrt{\beta_{1}}}{\sqrt{\pi}}+\right.\\\nonumber 
& \quad\left.+\left(1-2d^{2}\beta_{1}\right)e^{d^{2}\beta_{1}}\text{erfc}\left(d\sqrt{\beta_{1}}\right)\right],
\end{align}
where $\mathrm{erfc}\left(z\right)=1-\mathrm{erf}\left(z\right)$ is the complementary error function. 

Taking typical $\mathrm{hBN}$ parameters (effective mass $\mu=0.6\,m_{0}$, with $m_{0}$ the bare electron mass, and an effective dielectric constant $\epsilon_{r}=5.89$) and defining the effective
Bohr radius and the effective Rydberg as $a_{B}\equiv 4 \pi \epsilon_{0} \epsilon_{r} \hbar^{2} /\left(e^{2} \mu \right)=\frac{\hbar^{2}}{\mu k}=5.222\,\mathrm{\AA}$
and $\mathrm{Ry}\equiv \mu e^{4} /\left(32 \pi^{2} \epsilon_{0}^{2} \epsilon_{r}^{2} \hbar^{2}\right)=\frac{k}{2a_{B}}=234.1\,\mathrm{meV}$, the extrema of $\varepsilon_{i}$
for a separation bias $d$ in the typical range for this material
($5$ to $20\,\mathrm{\AA}$) are obtained and the variational energies
are plotted in Fig. \ref{fig:Optimal-variational-energy-1-1-1-1-1} (left panel). As expected, the binding energy increases as the distance between the layers decrease, due to the enhancement of the electrostatic interaction.

This procedure can be easily utilized for higher-energy states, with
the main drawback being the ever-increasing calculation time and complexity
of the Hamiltonian expectation value for the chosen \emph{ans{\"a}tze}. In this case, the expectation value of the Hamiltonian using the variational wave functions can always be computed numerically, if need be.

A more interesting application for this procedure is the repetition
of the calculations, now considering the Rytova-Keldysh potential.
Obtaining the variational energies with the same \emph{ans{\"a}tze},
the results for the two potentials can be more easily compared, both quantitatively and qualitatively.

\subsection{Intralayer Excitons in hBN: the Rytova-Keldysh potential}\label{subsec:Rytova-Keldysh-Potential}

\subsubsection{Introduction}

The purpose of this section is to discuss the formation of intralayer excitons in hBN as opposed to interlayer excitons described in the previous section. For intralayer excitons in 2D materials the suitable choice of electrostatic potential is the Rytova-Keldysh potential, as it provides a better description of screening in the 2D material. The Rytova-Keldysh potential in
polar coordinates can be written as \cite{rytova1967,1979JETPL..29..658K}
\begin{equation}
V_{RK}\left(r\right)=-\frac{k\pi}{2r_{0}}\left[H_{0}\left(\frac{r}{r_{0}}\right)-Y_{0}\left(\frac{r}{r_{0}}\right)\right]\label{eq:rytova-keldysh}
\end{equation}
where $k=e^{2}/\left(4\pi\epsilon_{0}\epsilon_{r}\right)$, $H_{0}\left(x\right)$
is the zeroth-order Struve function, $Y_{0}\left(x\right)$ is the zeroth-order
Bessel function of the second kind, and $r_{0}$ is a material-specific
screening length. The derivation of this potential, following the same approach as Cudazzo \emph{et al.} \cite{PhysRevB.84.085406}, is present in Appendix \ref{sec:deriv_keldysh}. Approximately, this potential behaves similarly
to \cite{PhysRevB.84.085406}
\begin{equation}
V_{RK}^{\prime}\left(r\right)\approx\frac{k}{r_{0}}\left[\log\left(\frac{r}{r+r_{0}}\right)+\left(\gamma-\log2\right)e^{-\frac{r}{r_{0}}}\right]
\end{equation}
which might help performing analytical calculations in the future.
Furthermore, when one takes the limit $r_{0}\rightarrow0$, the Rytova-Keldysh
potential (as well as its $\log+\exp$ approximation) reduces to the
usual Coulomb potential.

\subsubsection{Variational Energy}

In this section we use the same \emph{ans{\"a}tze} as that used in the 2D soft-Coulomb
problem. Taking the expectation values of the Hamiltonian with the
Rytova-Keldysh potential, we obtain 
\begin{align}
\epsilon_{0}\left(\beta_{0},r_{0}\right) & =\mathrm{Ry}\,a_{B}\left[\beta_{0}+\frac{\pi\,a_{B}\,G_{2,3}^{2,1}\left(\frac{1}{4r_{0}^{2}\beta_{0}}\left|\begin{array}{c}
0,-\frac{1}{2}\\
0,0,-\frac{1}{2}
\end{array}\right.\right)}{r_{0}}-\right.\nonumber\\
& \quad\left.-2\frac{\sqrt{\pi}\,a_{B}\,F\left(\frac{1}{2r_{0}\sqrt{\beta_{0}}}\right)}{r_{0}}\right],\label{eq:grasselli_rk}
\end{align}
and
\begin{equation}
\epsilon_{1}\left(\beta_{1},r_{0}\right)=\mathrm{Ry}\,a_{B}\left[2\beta_{1}-\frac{a_{B}\,G_{2,3}^{3,2}\left(\frac{1}{4r_{0}^{2}\beta_{1}}\left|\begin{array}{c}
-1,\frac{1}{2}\\
0,0,\frac{1}{2}
\end{array}\right.\right)}{\pi r_{0}}\right].
\end{equation}
Here, $F$ is the Dawson-F function and $G$ is the Meijer-G function,
a generalization of most special functions. The effective Bohr radius $a_B$ and the effective Rydberg $\mathrm{Ry}$ have the same numerical values as those defined previously. 

\begin{figure}
\centering

\includegraphics[scale=0.3]{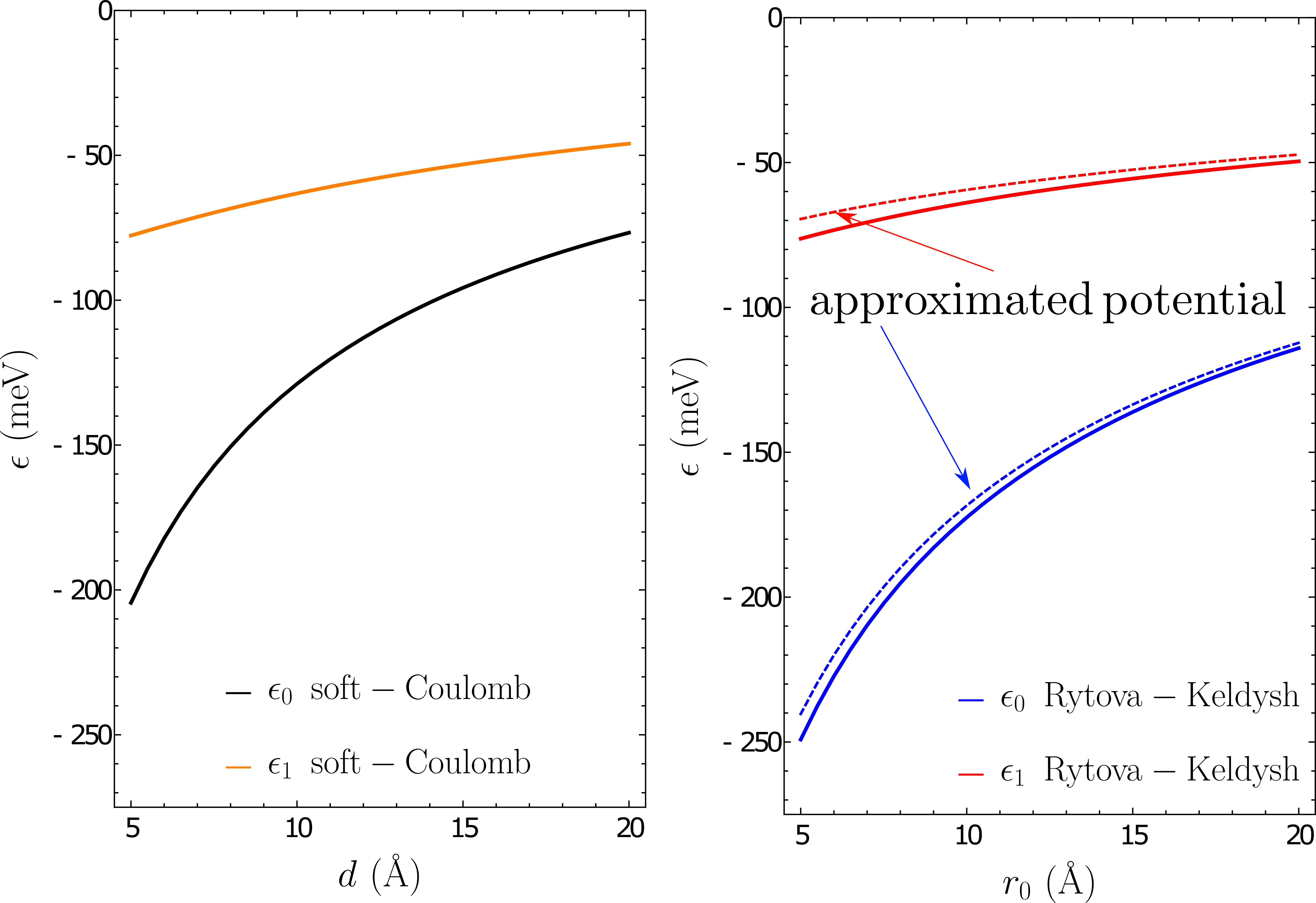}

\caption{\foreignlanguage{american}{(Left) Variational energy of both the ground and
the first excited state calculated for the soft-Coulomb potential
as a function of the separation bias $d\in\left[5,20\right]\,\mathrm{\AA}$. (Right) Comparison of the variational energy of both the ground and
the first excited state calculated for the Rytova-Keldysh (full lines) and the approximated form (dashed lines) as a function of the screening length $r_{0}\in\left[5,20\right]\,\mathrm{\AA}$.}}\label{fig:Optimal-variational-energy-1-1-1-1-1}
\end{figure}

Finding the extrema of these functions is significantly more difficult
than in the Soft-Coulomb problem. The spectrum in the range $r_{0}\in\left[5,20\right]\,\mathrm{\AA}$ is present in Figure
\ref{fig:Optimal-variational-energy-1-1-1-1-1} (right panel). 
As can be seen in this figure, there is a slight difference between
the $V_{RK}^{\prime}\left(r\right)$ approximation (dashed lines) and the exact Rytova-Keldysh potential (solid lines) which appears to increase as $r_{0}$ decreases.

As the computation time with the approximation of the Rytova-Keldysh is significantly slower, it seems preferable to just use the standard Rytova-Keldysh
potential when considering these specific \emph{ans{\"a}tze}.

\subsection{Different \emph{Ans\"{a}tze}}

Let us now consider a different \emph{ansatz},
given by
\begin{equation}
\phi\left(r\right)=\sqrt{\frac{2}{\pi}}\frac{1}{\beta}e^{-r/\beta},
\end{equation}
where $\beta$ is a variational parameter, which corresponds to the excitonic wave-function for a system with
the valence band completely full, the conduction band totally empty
and local screening \cite{Schmitt_Rink_1985}. The anisotropic formulation equivalent to this wave-function is given by \cite{Castellanos_Gomez_2014}
\begin{equation}
\phi\left(x,y\right)=\sqrt{\frac{2}{\pi\lambda_{1}\lambda_{2}}}e^{-\sqrt{\left(x/\lambda_{1}\right)^{2}+\left(y/\lambda_{2}\right)^{2}}},
\end{equation}
where both $\lambda_{1}$ and $\lambda_{2}$ are variational parameters, and was applied to excitons in phosphorene. The anisotropic Rytova-Keldysh potential, together with the anisotropic Stark shift and electroabsorption of excitons is discussed in Ref. \cite{kamban2020anisotropic}.

Choosing a different
\emph{ansatz }leads to a different form for the variational energy,
but the expectation value after minimization should be similar if the trial wave function is not too different from the exact groundstate. With
this \emph{ansatz}, the variational energy takes the form (where s-C stands for the soft-Coulomb potential and R-K stands for the Rytova-Keldysh potential)
\begin{align}
\epsilon_{\mathrm{s-C}}\left(\beta,r_{0}\right) & =\mathrm{Ry}\,a_{B}\left(\frac{a_{B}}{\beta^{2}}+\right.\nonumber\\
& \quad \left.+\frac{4\pi r_{0}\left[Y_{1}\left(\frac{2r_{0}}{\beta}\right)+H_{-1}\left(\frac{2r_{0}}{\beta}\right)\right]}{\beta^{2}}\right),\label{eq:rana_sc}
\end{align}
\begin{align}
\epsilon_{\mathrm{R-K}}\left(\beta,r_{0}\right) & =\mathrm{Ry}\,a_{B}\left(\frac{a_{B}}{\beta^{2}}-\frac{4\left(\beta-2r_{0}\right)}{\beta^{2}+4r_{0}^{2}}-\right.\nonumber\\
& \quad \left.-\frac{16r_{0}^{2}\left[\sinh^{-1}\left(\frac{2r_{0}}{\beta}\right)+\text{csch}^{-1}\left(\frac{2r_{0}}{\beta}\right)\right]}{\left(\beta^{2}+4r_{0}^{2}\right)^{3/2}}\right).\label{eq:rana_rk}
\end{align}
Equations \ref{eq:rana_sc}--\ref{eq:rana_rk} have very different
forms from Equations \ref{eq:grasselli_sc}--\ref{eq:grasselli_rk},
but the expectation values after minimization are quite similar, as the
variational method provides an upper bound to the (in this case) ground-state
energy. We note, however, that the shape of the wave functions differ.

\begin{figure}
\centering
\includegraphics[scale=0.3]{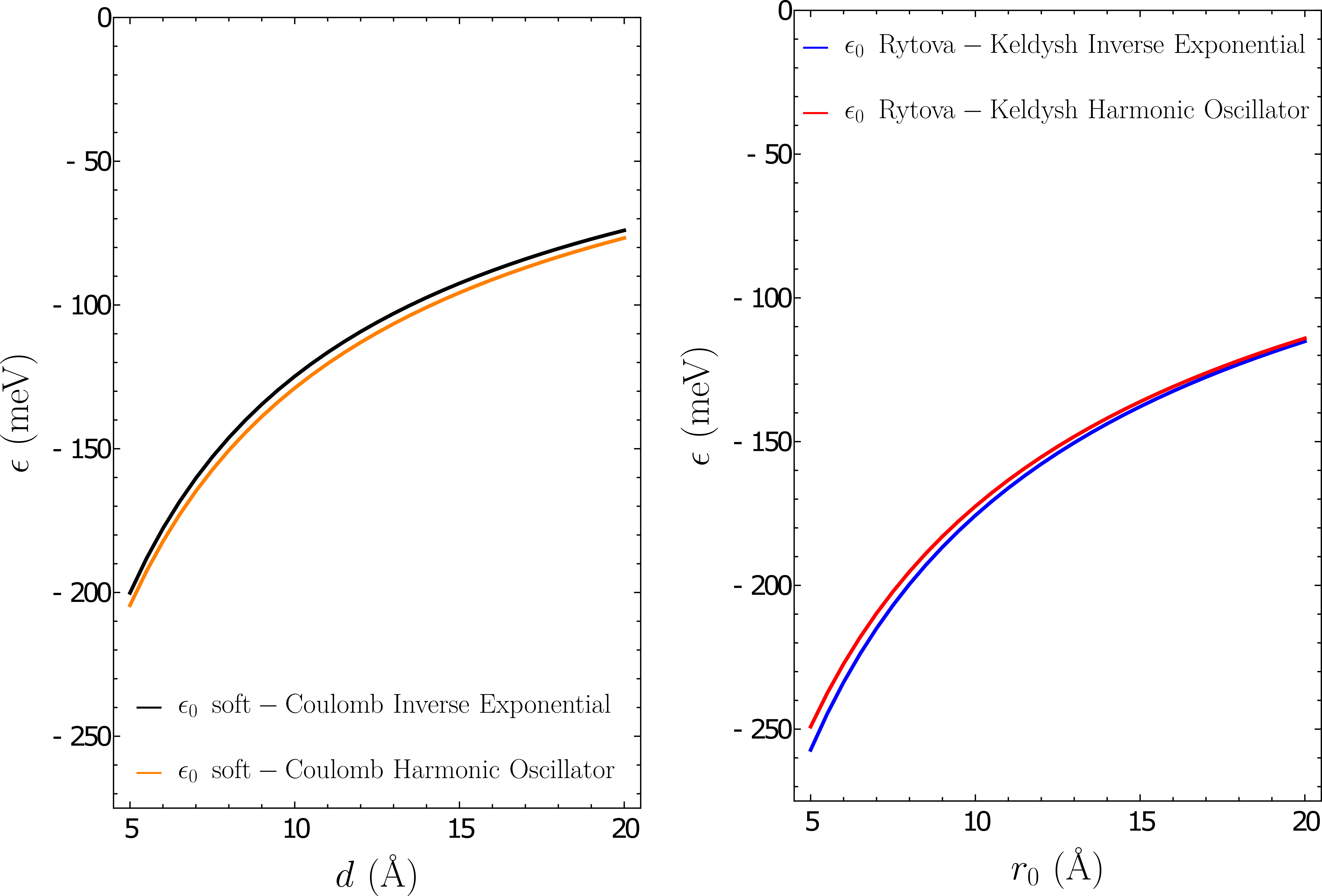}%
\caption{\foreignlanguage{american}{(Left) Comparison of the variational energy calculated for the soft-Coulomb potential with two different \emph{ans\"{a}tze}
as a function of the separation bias $d\in\left[5,20\right]\,\r{A}$}.
(Right) Comparison of the variational energy calculated for the Rytova-Keldysh potential with two different \emph{ans\"{a}tze}
as a function of the screening length $r_{0}\in\left[5,20\right]\,\r{A}$.}\label{fig:Optimal-variational-energy-ansatz}
\end{figure}

In Fig. \ref{fig:Optimal-variational-energy-ansatz} we compare the ground-state energy for the two different \emph{ans\"{a}tze} considered above and for the soft-Coulomb potential and the Rytova-Keldysh potential.
Looking at both panels of Fig. \ref{fig:Optimal-variational-energy-ansatz} 
we see that both \emph{ans\"{a}tze} give very similar ground-state energies.
In the case of the soft-Coulomb potential, the  harmonic oscillator \emph{ansatz}
gives a slightly smaller energy, whereas the Ritova-Keldysh potential the opposite 
happens. This inversion is related to the different behavior of the two potentials at the origin: the soft-Coulomb potential does not diverges at the origin whereas the 
Ritova-Keldysh does. Indeed, the soft-Coulomb potential can be approximated by a parabola at short distances and therefore a harmonic oscillator trial wave function should give a better description to the  true wave function of the ground state, as was effectively verified. In the case of a divergent potential at the origin,
an inverse exponential \emph{ansatz} should describe better the ground state wave function.

\section{The Wannier--Mott Exciton Model of a TMD Single-Layer}

The dielectric constant is generally large in semiconductors, which
in turn increases electric screening. This reduces the electron-hole
interaction, increasing the radius of the exciton. When this exciton
radius is greater than the lattice spacing, the exciton is a Wannier-Mott
exciton \cite{Wannier_1937,mott_basis_1949}. These are typically
found in semiconductors with small energy gaps and high dielectric
constants, as expected.

Transition-metal dichalcogenides, more specifically their 2D form, have emerged in the fields of electronics and
optoelectronics. One of their distinguishing features are the larger-than-usual
exciton (and trion) binding energies, almost an order of magnitude
larger compared to other bulk semiconductors. These large binding
energies imply that many-body interactions are fundamental to determining
and understanding the electronic and optoelectronic properties.

Using an equation of motion approach, Have \emph{et al.}\cite{Have_2019}
calculated excitonic properties of monolayer TMDs perturbed by an
external magnetic field. The obtained results are compared to both
the Wannier model for excitons, as well as recent experimental results.
A good agreement between the authors' calculated excitonic transition
energies and experimental data is observed, whilst being slightly
lower than those calculated via the Wannier exciton model. The authors
also show that the effects of the surrounding dielectric environment
in the magnetoexciton energy is minimal, as the changes in the exciton
energy and the exchange energy correction counteract each other. 

To better exemplify this exciton model, as well as compare the results
obtained from it against those from first--principles calculations,
we will briefly review the work of Zhang \emph{et al.}\cite{zhang_absorption_2014}.
In their work, the authors show that the traditional Wannier-Mott
exciton model with some modifications is able to appropriately describe
the exciton in 2D dichalcogenides. The modifications in question involve
taking into account accurate band structures of the conduction and
valence bands for large wave-vectors, incorporating phase-space cancellations
due to Pauli exclusion in doped materials, as well as taking into
account the finite thickness of dichalcogenide monolayers by considering
a wave-vector-dependent dielectric constant. With these modifications,
the authors study normally incident radiation, discussing the binding
energy and the absorption spectrum, both with and without taking into
account Pauli blocking.

\subsection{Exciton States}

Zhang \emph{et al.}\cite{zhang_absorption_2014} consider a system
where the initial state $\ket{\psi_{i}}$ of the semiconductor consists
of a completely filled valence band, as well as a conduction band
with an electron density $n_{e}$ in accordance with the Fermi-Dirac
distribution $f_{c}\left(\mathrm{k}\right)$. This initial state belongs to
a thermal ensemble and the average energy of the ground-state ($\ket{\psi_{i}}$) is
$E_{i}$ (i.e., $\bra{\psi_{i}}\hat{H}\ket{\psi_{i}}=E_{i}$).

Since only excitons with zero in-plane momentum are created by normally
incident radiation, an exciton state with zero in-plane momentum can
be constructed from the initial state as \cite{zhang_absorption_2014}
\begin{equation}
\ket{\psi_{\mathrm{ex}}}=\frac{1}{\sqrt{A}}\sum_{\mathbf{k}}\frac{\phi\left(\mathbf{k}\right)}{N_{\mathrm{ex}}\left(\mathbf{k}\right)}c_{\mathbf{k},\uparrow}^{\dagger}b_{\mathbf{k},\uparrow}\ket{\psi_{i}},\label{eq:wannier}
\end{equation}
where $c_{\mathbf{k},\uparrow},\,b_{\mathbf{k},\uparrow}$ are the destruction operators
for the spin-up conduction and valence-band states with momentum $\mathbf{k}$,
respectively, and $A$ is the area of the monolayer (normalization
factor). The normalization factor $N_{\mathrm{ex}}\left(\mathbf{k}\right)$
equals $\sqrt{1-f_{c}\left(\mathbf{k}\right)}$ and this exciton state is normalized
such that $\left\{ \braket{\psi_{\mathrm{ex}}}{\psi_{\mathrm{ex}}}\right\} _{\mathrm{th}}=1$,
where $\left\{ \ldots\right\} _{\mathrm{th}}$ represents averaging
with respect to the thermal ensemble). The presence of the Fermi-Dirac distribution in Eq. \ref{eq:wannier} allows the discussion of doped TMDs. In this context, it has been shown that the occuring blueshift of the binding energies depends significantly on the specific doping ($10\,\mathrm{meV}$ in electron-doped samples, while being absent in hole-doped samples) \cite{Chang_2019,Van_Tuan_2019}.  Note that only the $\tau=1$ valley, where the top most valence band is occupied by spin-up $\left(\sigma=1\right)$, has been considered in Eq. (\ref{eq:wannier}). Considering spin-down electrons will only multiply the final result by a degeneracy factor $g=2$, due to the contribution from the $\tau=-1$ valley, so no generality is lost when dealing only with spin-up electrons. 

The state defined in Eq. \ref{eq:wannier} is that of a Wannier
exciton, as we are assuming that Wannier exciton theory is
valid for 2D metal dichalcogenides, and is an eigenstate of the interacting
Hamiltonian \cite{mahan_many-particle-physics,Haug_2009}
\begin{align}
\hat{H} = & \sum_{\mathbf{k}} E_{c, \mathbf{k}} c_{\mathbf{k}}^{\dagger} c_{\mathbf{k}}+\sum_{\mathbf{k}} E_{v, \mathbf{k}} b_{\mathbf{k}}^{\dagger} b_{\mathbf{k}} + \nonumber \\
& +\frac{1}{2} \sum_{\mathbf{k}, \mathbf{k}^{\prime}, \mathbf{q} \neq 0} V_{\mathbf{q}}\left(c_{\mathbf{k}+\mathbf{q}}^{\dagger} c_{\mathbf{k}^{\prime}-\mathbf{q}}^{\dagger} c_{\mathbf{k}^{\prime}} c_{\mathbf{k}} + \right. \nonumber \\
& +\left. b_{\mathbf{k}+\mathbf{q}}^{\dagger} b_{\mathbf{k}^{\prime}-\mathbf{q}}^{\dagger} b_{\mathbf{k}^{\prime}} b_{\mathbf{k}}+2 c_{\mathbf{k}+\mathbf{q}}^{\dagger} b_{\mathbf{k}^{\prime}-\mathbf{q}}^{\dagger} b_{\mathbf{k}^{\prime}} c_{\mathbf{k}}\right)\label{WM_ham}
\end{align}
only when
$n_{e}=0$. As this is not the case, this state is taken as a variational
state, where the function $\phi\left(\mathbf{k}\right)$ can be varied to
minimize the expectation value $\left\{ \bra{\psi_{\mathrm{ex}}}\hat{H}\ket{\psi_{\mathrm{ex}}}\right\} _{\mathrm{th}}$. In Eq. \ref{WM_ham}, $E_{c, \mathbf{k}}$ and $E_{v, \mathbf{k}}$ are the conduction and valence band dispersion relations, respectively, while $V_{\mathbf{q}}$ is the interaction potential.

Manipulating the eigenvalue equation for this Hamiltonian (see Appendix \ref{sec:deriv_ham}), the Hermitian
eigenvalue equation is 
\begin{align}
\left[E_{\mathrm{ex}}-E_{i}\right]\phi\left(\mathbf{k}\right) =\left[\bar{E}_{c}\left(\mathbf{k}\right)-\bar{E}_{v}\left(\mathbf{k}\right)\right]\phi\left(\mathbf{k}\right)- & \nonumber\\
-\frac{\sqrt{1-f_{c}\left(\mathbf{k}\right)}}{A}\sum_{\mathbf{q}}V\left(\mathbf{q}\right)\phi\left(\mathbf{k}-\mathbf{q}\right)\sqrt{1-f_{c}\left(\mathbf{k}-\mathbf{q}\right)} &,\label{eq:variational_equation}
\end{align}
where $E_{\mathrm{ex}}$ is the average energy of the state $\ket{\psi_{\mathrm{ex}}}$
and $E_{\mathrm{ex}}-E_{i}=E_{g}-E_{\mathrm{exb}}$, where $E_{\mathrm{exb}}$
is the exciton binding energy and $E_{\mathrm{g}}$ is the energy gap. The eigenfunctions $\phi_{i}\left(\mathbf{k}\right)$
are defined as orthogonal and $\bar{E}_{c/v}\left(\mathbf{k}\right)$ are the
conduction/valence band dispersion relations, including exchange corrections.
The dielectric constant $\epsilon\left(q\right)$ is, in general,
frequency and wave-vector dependent \cite{Ouerdane_2011} and is given by (see the Appendix of Ref. \cite{zhang_absorption_2014})
\begin{equation}
\epsilon\left(q\right)=\epsilon_{2}\frac{1-\frac{\left(1-\epsilon_{2}/\epsilon_{1}\right)\left(1-\epsilon_{2}/\epsilon_{3}\right)}{\left(1+\epsilon_{2}/\epsilon_{1}\right)\left(1+\epsilon_{2}/\epsilon_{3}\right)}e^{-2qd}}{\left[1-\frac{\left(1-\epsilon_{2}/\epsilon_{1}\right)}{\left(1+\epsilon_{2}/\epsilon_{1}\right)}e^{-qd}\right]\left[1-\frac{\left(1-\epsilon_{2}/\epsilon_{3}\right)}{\left(1+\epsilon_{2}/\epsilon_{3}\right)}e^{-qd}\right]}
\label{eq:dielectric_const}
\end{equation}
for a $\mathrm{M}\mathrm{X}_{2}$ monolayer of thickness $d$ and
dielectric constant $\epsilon_{2}$ ``sandwiched'' between materials
with dielectric constants $\epsilon_{1},\,\epsilon_{3}$. Asymptotically,
this expression can be written as $\epsilon\left(q\right)=\frac{\epsilon_{1}+\epsilon_{3}}{2}$
in the $qd\ll1$ limit, and $\epsilon\left(q\right)=\epsilon_{2}$
for large wave vectors or very thick monolayers$\left(qd\gg1\right)$.
These limits match those expected considering the varying thickness
of the TMD monolayer.

The solutions to Eq. \ref{eq:variational_equation} represent bound excitons, as well as
electron-hole scattering states. These are, however, excluded since
their inclusion leads to modifications in the absorption spectrum
near the band edge far from the fundamental exciton line. The authors
assume the variational solution
\begin{equation}
\begin{array}{ccc}
\phi\left(\mathbf{k}\right)=\frac{\sqrt{8\pi}a}{\left[1+\left(ka\right)^{2}\right]^{3/2}} & \implies & \phi\left(\mathbf{r}\right)=\sqrt{\frac{2}{\pi}}\frac{1}{a}e^{-r/a}\end{array},\label{eq:variational-functs_wannier}
\end{equation}
which is the exact exciton wave function for the Coulomb potential when $n_{e}=0$ and $\epsilon\left(q\right)$
is independent of $q$ \cite{Schmitt_Rink_1985} (i.e., local screening).
Varying the radius parameter $a$, the eigenvalue $E_{\mathrm{ex}}-E_{i}$
can be estimated.

\subsection{Optical Conductivity and Absorption Spectrum\label{sec:exciton_radius}}

The electronic and optical properties of a TMD are determined by the low energy Hamiltonian of the system. In the case we are considering, this Hamiltonian is valid around the $\mathcal{K}$ and $\mathcal{K}^{\prime}$ points of the hexagonal Brillouin zone, where the band gap is located. Effectively, the low energy Hamiltonian is similar to a Dirac Hamiltonian in 2D, with an additional term due to spin-orbit coupling, induced by the heavy metal atoms. 
The Hamiltonian near the $\mathcal{K},\,\mathcal{K}^{\prime}$ points is given by 
\begin{equation}
H_{0}=\left[\begin{array}{cc}
\Delta/2 & \hbar v k_{-} \\
\hbar v k_{+}, & -\Delta / 2+\lambda \tau \sigma
\end{array}\right],
\end{equation}
where $\Delta$ is related to the material band gap, $\sigma=\pm 1$ stands for the electron spin, $\tau=\pm 1$ is the valley index $\left(\mathcal{K},\,\mathcal{K}^{\prime}\right)$, $2\lambda$ is the splitting of the valence band due to spin-orbit coupling, $k_{\pm}=\tau k_{x} \pm i k_{y}$ (the wave vectors are measured from the $\mathcal{K}\,\left(\mathcal{K}^{\prime}\right)$ points), and the velocity parameter $v\approx5-6\times10^{5}\,\mathrm{m/s}$ is related to the coupling between the orbitals of neighboring $\mathrm{M}$ atoms. 

Zhang \emph{et al.}\cite{zhang_absorption_2014} assume linearly polarized
light of frequency $\omega$ and intensity $I_{0}$ incident normally
on the $\mathrm{M}\mathrm{X}_{2}$ monolayer. The electric field is
obtained directly from Maxwell's equations and, by definition of the
Poynting vector (time averaged for a linearly polarized electromagnetic
plane wave), we have
\begin{equation}
I_{0}=\frac{1}{2\eta_{0}}\left|E_{0}\right|^{2}=\frac{A_{0}^{2}\omega^{2}}{2\eta_{0}},
\end{equation}
where $\eta_{0}$ is the free-space impedance and $A_{0}$ is the
intensity of the vector potential in the monolayer plane. 

The interaction between spin-up electrons in the valley $\tau=+1$
and light is given by the Hamiltonian \cite{have_excitonic_2019,Peres_2010}
\begin{align}
H_{int}\left(t\right) & =H_{+}e^{-i\omega t}+\mathrm{h.c.}\nonumber\\&=\frac{eA_{0}}{2m_{0}}\sum_{\mathbf{k}}\vec{P}_{cv}\left(\mathbf{k}\right)\cdot\hat{n}e^{-i\omega t}c_{\mathbf{k},\uparrow}^{\dagger}b_{\mathbf{k},\uparrow}+\mathrm{h.c.},\label{eq:H_int}
\end{align}
where $\hat{n}$ is the polarization vector (in the plane of the monolayer),
$c_{\mathbf{k},\uparrow},\,b_{\mathbf{k},\uparrow}$ are, respectively, the destruction
operators for the spin-up conduction and valence-band states with
momentum $\mathbf{k}$, $m_{0}$ is the free-electron mass and $\vec{P}_{cv}\left(\mathbf{k}\right)$
is the momentum matrix element, taken between the conduction and valence
bands.

From Fermi's golden rule, the rate at which excitons are generated
by the absorption of light assuming finite broadening $\Gamma_{\mathrm{ex}}$ is given by
\cite{Glutsch_2004}
\begin{align}
R_{\mathrm{ex}} &=\frac{2\pi}{\hbar}\frac{1}{A}\left\{\left|\bra{\psi_{\mathrm{ex}}}\hat{H}_{+}\ket{\psi_{\mathrm{i}}}\right|^{2}\right\}_{\mathrm{th}}\times\nonumber\\
 & \quad\times\frac{\hbar\Gamma_{\mathrm{ex}}/\pi}{\left(\hbar\omega_{0}-\hbar\omega\right)^{2}+\left( \hbar\Gamma_{\mathrm{ex}}\right) ^{2}},\label{eq:fermi_rule}
\end{align}
where $\hbar\omega_{0}=E_{\mathrm{ex}}-E_{\mathrm{i}}$. Directly replacing $H_{+}$ from Eq. \ref{eq:H_int} and expanding
the modulus squared, we obtain 
\begin{align}
\frac{1}{A}\left\{ \left|\bra{\psi_{\mathrm{ex}}}\hat{H}_{+}\ket{\psi_{\mathrm{i}}}\right|^{2}\right\} _{\mathrm{th}} & =\left(\frac{eA_{0}}{2m_{0}}\right)^{2}\left|\chi_{\mathrm{ex}}\left(\mathrm{r}=0\right)\right|^{2}
\end{align}
where 
\begin{equation}
\chi_{\mathrm{ex}}\left(\mathbf{r}\right)=\int\frac{d^{2}\mathbf{k}}{\left(2\pi\right)^{2}}\phi\left(\mathbf{k}\right)\sqrt{1-f_{c}\left(\mathbf{k}\right)}\left[\vec{P}_{cv}\left(\mathbf{k}\right)\cdot\hat{n}\right]e^{i\mathbf{k}\cdot\mathbf{r}}\label{eq:chi_ex}.
\end{equation}

Different processes, such as scattering and inhomogeneous broadening
are expected to contribute to the absorption width $\Gamma_{\mathrm{ex}}$.
Equation \ref{eq:chi_ex} incorporates the effects (reduction in exciton oscillator
strength) due to Pauli blocking. The total energy absorption rate
from both valleys in the Brillouin zone $\left(\mathcal{K},\,\mathcal{K}^{\prime}\right)$
is given by $2\hbar\omega R_{\mathrm{ex}}$, and can be written in
terms of the exciton contribution to the optical conductivity, which
is 
\begin{align}
\mathrm{Re}\left\{ \sigma_{\mathrm{ex}}\left(\omega\right)\right\}  & =\frac{2\hbar\omega}{\eta_{0}I_{0}}R_{ex}\nonumber\\& =\frac{e^{2}}{4\hbar}\left\{ \frac{8\hbar}{m_{0}^{2}\omega_{0}}\left|\chi_{\mathrm{ex}}\left(\mathbf{r}=0\right)\right|^{2}\times\right.\nonumber\\
& \quad\times\left.\frac{\hbar\Gamma_{\mathrm{ex}}}{\left(\hbar\omega_{0}-\hbar\omega\right)^{2}+\left( \hbar\Gamma_{\mathrm{ex}}\right) ^{2}}\right\} .\label{eq:W-M_exciton_model}
\end{align}
Once the real part of the conductivity has been determined, the imaginary part follows from the Kramers-Kronig relations. The absorption spectrum for normally incident light can be obtained
from the optical conductivity \cite{Haug_2009} as 
\begin{equation}
A\left(\omega\right)\approx\frac{2\mathrm{Re}\left\{ \sigma_{\mathrm{ex}}\left(\omega\right)\right\} \eta_{0}}{1+n_{sub}},
\end{equation}
where $n_{sub}$ is the refractive index of the substrate (in the
case discussed by the authors, quartz).

Ignoring the effects of Pauli blocking and considering the wave-vector-independent
expression for the momentum matrix element in Eqs. \ref{eq:fermi_rule} and \ref{eq:W-M_exciton_model}, the exciton optical conductivity
is given by
\begin{align}
\mathrm{Re}\left\{ \sigma_{\mathrm{ex}}\left(\omega\right)\right\}  & =\frac{2e^{2}v^{2}}{\omega_{0}}\frac{2}{\pi a^{2}}\frac{\hbar\Gamma_{\mathrm{ex}}}{\left(\hbar\omega_{0}-\hbar\omega\right)^{2}+\left( \hbar\Gamma_{\mathrm{ex}}\right) ^{2}}
\end{align}
and the product $\left.A\left(\omega\right)\right|_{max}2\hbar\Gamma_{ex}$ by 
\begin{align}
\left.A\left(\omega\right)\right|_{max}2\hbar\Gamma_{ex} & =\frac{16\eta_{0}}{1+n_{sub}}\frac{e^{2}v^{2}}{\pi\omega_{0}}\frac{1}{a^{2}} .
\end{align}
By comparison with the extracted peaks, the exciton radius $a$ is
found to be around $\sim16.8\,\text{\AA}$. This radius is much larger than the lattice spacing, thus justifying the validity of the Wannier model. This approach is however too simplistic. A more realistic approach takes trigonal warping into account by adding more terms to $H_0$.

Taking into account the wave-vector dependence of the momentum matrix
elements (due to trigonal warping), as well as Pauli blocking, the absorption spectrum reads 
\begin{equation}
\left.A\left(\omega\right)\right|_{max}2\hbar\Gamma_{ex}=\frac{16\eta_{0}}{1+n_{sub}}\frac{e^{2}}{2m_{0}^{2}\omega_{0}}\left|\chi_{\mathrm{ex}}\left(\mathbf{r}=0\right)\right|^{2},
\end{equation}
from which the exciton radius can be extracted given the electron
density. For carrier densities around $2-4\times10^{12}\,\mathrm{cm}^{-2}$,
Zhang \emph{et al.} obtain the exciton radius in the $9.3-8.5\,\text{\AA}$
range, in agreement with first-principles theoretical estimates\cite{zhang_absorption_2014}.

In Ref. \cite{zhang_absorption_2014}, the absorption spectrum is discussed and compared to the measured
spectrum of $\mathrm{Mo}\mathrm{S}_{2}$ at different temperatures.
Comparing the calculated spectra against the extracted contributions
for the specific materials and excitons in question (at both $T=5\,K$
and $T=90\,K$), Zhang \emph{et al.} obtain a good fit between the two curves.
Furthermore, the individual contributions from both excitons and trions
(not discussed in this colloquium) appear to be reliably extractable using
this fitting procedure.

\subsection{Binding Energy\label{sec:binding_energy}}

As the obtained expression for the absorption spectrum has a free
parameter, the exciton radius, the resulting curves are superposed
against the experimental fits for various temperatures. By matching the
peaks, the value for the exciton radius is extracted. The only free parameter
in the variational eigenvalue equation becomes the momentum-- and frequency--dependent dielectric constant (Eq. \ref{eq:dielectric_const}),
which can be obtained by applying the variational principle to Eq.\ref{eq:variational_equation}
using the obtained variational functions in Eq.\ref{eq:variational-functs_wannier},
and matching the resulting exciton radii with the values obtained
from the absorption peaks. Considering that the top--bottom Sulfur distance in $\mathrm{MoS}_{2}$ is
$\sim3.17\,\text{\AA}$ (the effective monolayer thickness is taken
as $d\sim6\,\text{\AA}$ \cite{zhang_absorption_2014}) and that we have a quartz substrate ($\epsilon_{3}=4$)
and $\epsilon_{1}$ as free space ($=1$), the value of $\epsilon_{2}$
for which the exciton radii matches the measured values is $\epsilon_{2}\sim12$.
While rather large, this value matches well with the theoretical estimates
for the bulk $\mathrm{MoS}_{2}$ dielectric constant. Furthermore,
following Berkelbach \emph{et al.}\cite{Berkelbach_2013} and Cudazzo
\emph{et al.}\cite{Cudazzo_2011}, one finds the screening length parameter
$r_{0}$. The resulting value, $36\,\text{\AA}$, is in excellent
agreement with first-principles calculations by Berkelbach \emph{et
al}\cite{Berkelbach_2013} ($30-40\,\text{\AA}$). The variational
binding energy for the extracted values is in the $0.28-0.33\,\mathrm{eV}$
range, in good agreement with the first-principles calculations \cite{Berkelbach_2013,Qiu_2013}.

\section{Interlayer Excitons in a TMD: The Harmonic Potential Approximation}

Many authors discuss the optical and electronic properties of excitons
in TMDs and vdW heterostructures from a numerical point
of view, usually by directly integrating the Schr\"{o}dinger equation
with either the Rytova-Keldysh potential (via finite elements, for
example). Brunetti \emph{et al.}\cite{brunetti_optical_2018} verify
the accuracy of the numerical results for large interlayer separations
by approximating the electron-hole interaction as an harmonic potential.
As the Schr\"{o}dinger equation is solvable exactly for this interaction,
both the wave function and the energy eigenvalue can be written in
close form and used to verify the validity of numerically calculated
physical quantities such as the (peak) absorption coefficient and
the exciton binding energy.

Among the most recent numerical studies, we mention specifically those
by Avalos-Ovando \emph{et al.}\cite{avalos-ovando_lateral_2019}, Scharf
\emph{et al.}\cite{Scharf_2019}, Van der Donck \emph{et al.}\cite{van_der_donck_excitons_2017}
and Brunetti \emph{et al.}\cite{brunetti_optical_2018}. We will now
briefly discuss the results by Brunetti \emph{et al.}.

\subsection{The Harmonic Potential Approximation}

In the case of interlayer excitons, the modified Rytova-Keldysh potential (defined ahead) shows a minimum at $r=0$ and has a finite value at that point. This happens because interlayer  excitons are spatially separated by a distance $d$.
Then, we can expand the modified potential around $r=0$ up to second order, leading to the harmonic approximation.
In this work \cite{brunetti_optical_2018}, the authors discuss the
optical absorption of interlayer excitons in a vdW heterostructure.
Starting from the oscillator strength for a generic transition
\begin{equation}
f_{i\rightarrow f}=\frac{2\mu\omega_{i\rightarrow f}\left|\bra fx\ket i\right|^{2}}{\hbar},
\end{equation}
the imaginary part of the electric susceptibility is given by \cite{snoke2009solid}
\begin{equation}
\mathrm{Im}\left\{ \chi\left(\omega\right)\right\} =-\frac{\pi e^{2}}{2\varepsilon_{0}\mu\omega_{0}}\frac{n_{0}}{2h}f_{0}\frac{\Gamma_{\mathrm{ex}}/2}{\left(\omega_{0}^{2}-\omega^{2}\right)^{2}+\left(\Gamma_{\mathrm{ex}}/2\right)^{2}},
\end{equation}
where $n_{0}$ is the 2D concentration of excitons in the heterostructure,
$h$ is the thickness of one TMD monolayer, $\Gamma_{\mathrm{ex}}$ the homogeneous
line-broadening (with physical origin in exciton-phonon interactions),
$f_{0}$ represents the oscillator strength for a generic transition, $\omega_0$ refers to the respective Bohr angular frequency of the oscillator strength $f_{0}$, and $\mu$ is the exciton reduced mass. Knowing this, the optical
absorption coefficient is given by \cite{jackson1999classical,Haug_2009} ($c$ is the speed of light) 
\begin{equation}
\alpha\left(\omega\right)=-\frac{\omega}{n\left(\omega\right)c}\mathrm{Im}\left\{ \chi\left(\omega\right)\right\} ,
\end{equation}
which can be further simplified assuming the environment interacts
weakly with photons in the frequency range of the corresponding optical
transition, approximating the refractive index of the environment
as $n\left(\omega\right)\approx\sqrt{\varepsilon}$, with $\varepsilon$
being the static dielectric constant of the environment. At the peak, the absorption
coefficient is given by 
\begin{equation}
\alpha\left(\omega=\omega_{0}\right)=\frac{\pi e^{2}}{2\varepsilon_{0}\mu\sqrt{\varepsilon}c}\frac{n_{0}}{2h}f_{0}\frac{2}{\Gamma_{\mathrm{ex}}},
\end{equation}
which is defined by the oscillator strength, and depends on a series
of material-specific parameters.

The interlayer modified Rytova-Keldysh potential reads
\begin{equation}
V_{RK}\left(r\right)=-\frac{\pi k}{2 r_{0}}\left[H_{0}\left(\frac{\sqrt{r^{2}+d^{2}}}{r_{0}^{2}}\right)-Y_{0}\left(\frac{\sqrt{r^{2}+d^{2}}}{r_{0}^{2}}\right)\right],
\end{equation}
with $k=e^{2}/4\pi\epsilon_{0}\kappa$ and $\kappa=\frac{\epsilon_{1}+\epsilon{2}}{2}$ describes the surrounding dielectric medium, is approximated by a harmonic potential by considering the interlayer
separation $d$ much larger than the in-plane gyration
radius $\left(r_{X}=\sqrt{\left\langle r^{2}\right\rangle }\right.$, that is, the average of $r^2$ taken with the exciton wave function).
This approximation is given by \cite{Berman-2017,brunetti_optical_2018}
\begin{equation}
V\left(r\right)\approx-V_{0}+\gamma r^{2}, 
\end{equation}
with $r_{0}$ the dielectric screening length and 
\begin{align}
V_{0} & =\frac{\pi k}{2 r_{0}}\left[H_{0}\left(\frac{d}{r_{0}}\right)-Y_{0}\left(\frac{d}{r_{0}}\right)\right]\nonumber\\ \gamma & =-\frac{\pi k}{4 r_{0}^{2}d}\left[H_{-1}\left(\frac{d}{r_{0}}\right)-Y_{-1}\left(\frac{d}{r_{0}}\right)\right].
\end{align}

Considering this approximation, the Schr\"{o}dinger equation can be solved
directly in polar coordinates as \cite{Edery_2018,cohen-tannoudji_quantum_2019}
\begin{align}
\Psi\left(\eta,\varphi\right)&=A_{nl}\left(\frac{\hbar}{\mu \omega}\right)^{\vert l\vert/2}\eta^{\vert l\vert /2}e^{-\eta/2}L_n^{\vert l\vert }(\eta)\frac{e^{i l \varphi}}{\sqrt{2\pi}},
\end{align}
where $\eta=\mu\omega r^2/\hbar$, $\omega=\sqrt{2\gamma/\mu}$, 
$n=0,1,2,\ldots$ and $l=-n,-n+1,\ldots,n-1,n$ are the usual
principal and angular momentum quantum numbers, and $L_{b}^{a}\left(x\right)$ is
the associate Laguerre polynomial of degree $b$ and order $a$, 
and
\begin{equation}
A_{nl}=\left(
\frac{\mu \omega}{\hbar}
\right)^{(\vert l\vert +1)/2}\sqrt{\frac{2n!}{\Gamma(n+\vert l\vert +1)}},
\end{equation}
with $\Gamma(x)$ the gamma-function.
The energy eigenvalues
present the usual form for the isotropic harmonic oscillator, given by 
\begin{equation}
E_{n}=\hbar\left(\frac{2\gamma}{\mu}\right)^{1/2}\left(2n+\vert l\vert +1\right)-V_{0}.
\end{equation}

This approximation becomes more accurate the more layers are included
in the heterostructure (as expected by the assumption of large interlayer
separation), and it is verified by comparison with the results from
direct numerical integration (through finite elements methods), as
well as those obtained experimentally \cite{Horng_2018,Rahaman2019} or via DFT \cite{Kyl_np__2015}.

This approach agrees with the numerical results obtained by Brunetti
\emph{et al.}\cite{brunetti_optical_2018}, validating the numerical
approach utilized. The eigenenergies and eigenfunctions for interlayer
$\left(d\neq 0\right)$ excitons are calculated for a range of material-specific
parameters consistent with those from different TMDs. Comparing against
existing DFT results \cite{Kyl_np__2015}, an agreement to better
than $1\%$ is obtained. As such, eigenvalues and optical properties
for interlayer excitons can be calculated for different ranges of input
parameters consistent with different TMD/h-BN/TMD heterostructures.

In a recent experimental work \cite{Horng_2018}, spatially-indirect
excitons were observed in a single $\mathrm{Mo}\mathrm{Se}_{2}$ crystal.
In bilayer $\mathrm{Mo}\mathrm{Se}_{2}$ encapsulated by h-BN, spatially-indirect
excitons are reported to have a binding energy of $153\,\mathrm{meV}$.
For the material-dependent parameters provided by \cite{Horng_2018}, and using the oscillator strength $f_0$ for the transition between the $n=1\rightarrow 2,\,l=0\rightarrow 1$ eigenstates  \cite{brunetti_optical_2018}, Brunetti's calculations give a binding energy of $132-140\,\mathrm{meV}$, within $\approx10\%$
of the reported value.

\section{Conclusions and Outlook}

The objective of this colloquium was to present an analysis of excitons in 2D materials and vdW heterostructures using the variational method. 
We have focused our attention in transition metal dichalcogenides that have a direct band gap at the $\cal K$ and $ \cal K^\prime$ points of the Brillouin zone. In these systems, the electronic and optical properties are described by an effective theory based on the 2D Dirac equation. However, other 2D systems exist presenting indirect band gap between the $\Gamma$ and the M points, such as ZrS$_2$ and HfS$_2$ \cite{Lau_2019}.

We have first analyzed variational methods for obtaining the binding energy of vdW
excitons whose interaction is described by the soft-Coulomb potential. This potential
includes a material-dependent parameter which eliminates the divergence
of the Coulomb potential at the origin. This material dependent parameter  appears  in the electrostatic interaction when
 an electron is in a given layer and the hole is in a different one,  a distance
$d$ apart from each other. This can happen both in bulk single crystals, due to the layered nature of these materials, or in heterogeneous systems made by staking different 2D materials on top of each other. In simple systems, analytical solutions for soft-Coulomb potential are known, thus allowing for the comparison of the variational approach
with exact analytical results.

After presenting the variational method in the context of the soft-Coulomb potential, we 
considered the Rytova-Keldysh potential. This interaction describes better
the scenario of a monolayer transition-metal dichalcogenide both suspended, supported by a substrate, or  encapsulated by low-dielectric materials. This
potential is mathematically more complex than the soft-Coulomb one. It can,
however, be closely approximated by a logarithmic and an inverse-exponential
term, which may allow the calculation of the binding energies analytically.

Although we have not considered van der Waals  excitons in 2D quantum dots (triangular and hexagonal,
as there are no circular TMD systems) the introduced methodology is appropriate to the study of these systems, as wave functions with the correct symmetry can 
be easily constructed using, for example, group-symmetry arguments.

For connecting with experimental results, both the absorption spectrum
and the polarizability of  excitons was discussed. These two
properties are extremely important in optics since they govern the system's
response to optical impulses. The Stark effect and ionization rate
of van der Waals heterostructures have also been considered, as these effects can considerably
enhance the photodetection response of photodetectors.

This colloquium does not exhaust all the possibilities  of the variational method, with trions and bi-excitons physical properties also being amenable  to this type of treatment. Also, the inclusion of a magnetic field causes no difficulties to the method.

\section*{Acknowledgments}
The authors thank Eduardo Castro and Jo\~{a}o Lopes dos Santos for comments on a preliminary version of this work, and Bruno Amorim for outlining the derivation in the Appendix \ref{sec:appendix_BA}.
N.M.R.P. acknowledges support from the European Commission through the project Graphene-Driven Revolutions in ICT and Beyond (Ref. No. 881603 -- core 3), and the Portuguese Foundation for Science and Technology (FCT) in the framework of the Strategic Financing UID/FIS/04650/2019. In addition, funding from the projects POCI-01-0145-FEDER-028114, and POCI-01-0145-FEDER-029265, and PTDC/NAN-OPT/29265/2017, and POCI-01-0145-FEDER-02888 is acknowledged.
%\section{Authors contributions}
%All the authors were involved in the preparation of the manuscript.
%All the authors have read and approved the final manuscript.
%

\appendix
\section{Exciton dipole matrix element in the Wannier model: two-band system\label{sec:appendix_BA}}

In this Appendix we show that under specific conditions it is the Fourier transform to
real space, at the origin of the relative coordinate, of the exciton
wave function in momentum space that determines the dipole matrix
element associated with the optical transitions referred in the main
text. Furthermore, we will see that this result is indeed an approximation
and does not hold in general. This is specially true when the wave
function of the electron and the hole are defined by momentum-dependent
spinors. Let us consider an exciton with center-of-mass momentum $\mathbf{Q}$.
The excitonic wave function reads
\begin{equation}
\vert\Psi_{X,\mathbf{Q}}\rangle=\frac{1}{\sqrt{A}}\sum_{\mathbf{k}}\phi_{cv}(\mathbf{k})\vert\psi_{c,\mathbf{k}-\mathbf{Q}/2}^{*},\psi_{v,\mathbf{k}+\mathbf{Q}/2}\rangle,
\end{equation}
where $\phi_{cv}(\mathbf{k})$ is the exciton wave function is momentum
space and $\vert\Psi_{X,\mathbf{Q}}\rangle$ represents a sum of Slater
determinants with excitons of momentum $\mathbf{k}$. Therefore, we
obtain for the dipole matrix element between the ground state, $\vert0\rangle$
(a Slater determinant), of the TMD and its excited state the result \cite{0080168469}
\begin{align}
\langle0\vert\mathbf{\sum_{i}\mathbf{r}_{i}}\vert\Psi_{X,\mathbf{Q}}\rangle&\equiv\mathbf{d}_{0,X}=\frac{1}{\sqrt{A}}\sum_{\mathbf{k}}\phi_{cv}(\mathbf{k})\times\\\nonumber
&\quad\times\int d\mathbf{r}\psi_{c,\mathbf{k}-\mathbf{Q}/2}^{*}(\mathbf{r})\mathbf{r}\psi_{v,\mathbf{k}+\mathbf{Q}/2}(\mathbf{r}),
\end{align}
where $\mathbf{r}$ is the relative coordinate and $\sum_{i}\mathbf{r}_{i}$
is a sum over the position of all the electrons in the solid. Defining
\begin{equation}
\mathbf{d}_{c,\mathbf{k}-\mathbf{Q}/2;v,\mathbf{k}+\mathbf{Q}/2}\equiv\int d\mathbf{r}\psi_{c,\mathbf{k}-\mathbf{Q}/2}^{*}(\mathbf{r})\mathbf{r}\psi_{v,\mathbf{k}+\mathbf{Q}/2}(\mathbf{r}),
\end{equation}
we obtain
\begin{equation}
\mathbf{d}_{0,X}=\frac{1}{\sqrt{A}}\sum_{\mathbf{k}}\phi_{cv}(\mathbf{k})\mathbf{d}_{c,\mathbf{k}-\mathbf{Q}/2;v,\mathbf{k}+\mathbf{Q}/2}.
\end{equation}
Now, consider excitons with zero center of mass momentum, $\mathbf{Q}=0$,
described by plane waves. Assuming that the momentum dipole matrix
element is weakly dependent on $\mathbf{k}$ in this case, we can
approximate
\begin{equation}
\mathbf{d}_{c,\mathbf{k}-\mathbf{Q}/2;v,\mathbf{k}+\mathbf{Q}/2}\simeq\frac{1}{A}\mathbf{d}_{c;v},
\end{equation}
such that the exciton dipole moment is given by
\begin{align}
\mathbf{d}_{0,X} & \simeq\frac{1}{\sqrt{A}}\frac{1}{A}\sum_{\mathbf{k}}\phi(\mathbf{k})\mathbf{d}_{c;v}\nonumber \\
 & =\frac{1}{\sqrt{A}}\chi_{{\rm ex}}(\mathbf{r}=0)\mathbf{d}_{c;v}.\label{eq:dipole_plane_wave}
\end{align}
Therefore, we see that it is the Fourier transform to real space (at
the origin), $\chi_{{\rm ex}}(\mathbf{r}=0)$, of the exciton wave
function in momentum space that determines the optical response of
the semiconductor. We note that if the electron and hole wave function
cannot be taken as plane waves then extra momentum-dependent factors
appear and we cannot obtain the simple result given in Eq. (\ref{eq:dipole_plane_wave}).
In this case, other states, in addition to the $s-$state (the only
one finite at the origin) of the exciton wave function, contribute
to the optical response.

\section{Derivation of the Rytova-Keldysh potential\label{sec:deriv_keldysh}}

In this Appendix, we perform a derivation of a screened potential based on the assumption
that the charge fluctuation is proportional to the Laplacian of the
potential evaluated in the plane of the 2D material surrounded by vacuum \cite{PhysRevB.84.085406}.
Such an assumption comes from the following considerations: the induced
charge density $\delta n_{2D}\left(\mathbf{r}_{\parallel}\right)$ due to a point charge in the system is given by the
2D polarization in the usual way 
\begin{equation}
\delta n_{2D}\left(\mathbf{r}_{\parallel}\right)=-\nabla\cdot\mathbf{P}_{2D},
\end{equation}
where the three-dimensional position vector is given by $\mathbf{r}=(\mathbf{r}_\parallel,z)$, and $\delta n_{2D}$ has units of charge per unit area.
The polarization itself is proportional to the total electric field
\begin{equation}
\mathbf{P}_{2D}=-\epsilon_0\chi_{2D}\nabla V\left(\mathbf{r}_{\parallel},z=0\right),
\end{equation}
with $\chi_{2D}$ having dimensions of length. Therefore,

\begin{equation}
\delta n_{2D}\left(\mathbf{r}_{\parallel}\right)=\epsilon_0\chi_{2D}\nabla^{2}V\left(\mathbf{r}_{\parallel},z=0\right).
\end{equation}
Let us write Poisson's equation as:
\begin{equation}
\nabla^{2}V\left(\mathbf{r}\right)=-\frac{e}{\epsilon_{0}}\left[n_{2D,+}+n\left(\mathbf{r}\right)\right],
\end{equation}
where $n_{2D,+}$ is the background positive charge density due to the atomic nuclei.
We now write the electronic density as $n\left(\mathbf{r}\right)=-n_{2D,-}+\delta\left(\mathbf{r}\right)+\delta\left(z\right)\delta \sigma\left(\mathbf{r}_{\parallel}\right)$,
where $n_{2D,-}$ is the neutralizing density of negative charge,
$\delta\left(\mathbf{r}\right)$ represents the density of a localized charge
and $\delta\left(z\right)\delta \sigma\left(\mathbf{r}_{\parallel}\right)$ is the induced charge
density. With these definitions Poisson's equation
reads

\begin{align}
\nabla^{2}V\left(\mathbf{r}\right) & = -\frac{e}{\epsilon_{0}}\left[\delta\left(\mathbf{r}\right)+\delta\left(z\right)\delta \sigma\left(\mathbf{r}_{\parallel}\right)\right]\nonumber\\
& =-\frac{e}{\epsilon_{0}}\delta\left(\mathbf{r}\right)-\delta\left(z\right)\chi_{2D}\nabla^{2}V\left(\mathbf{r}_{\parallel},0\right),
\end{align}
where $e\delta \sigma (\mathbf{r}_\parallel)=\delta n_{2D} (\mathbf{r}_\parallel)$. Fourier transforming the previous
equation we obtain
\begin{align}
\nabla^{2} & \int\frac{d\mathbf{k}}{\left(2\text{\ensuremath{\pi}}\right)^{3}}e^{i\mathbf{k}\cdot\mathbf{r}}V\left(\mathbf{k}\right)  = -\frac{e}{\epsilon_{0}}\int\frac{d\mathbf{k}}{\left(2\text{\ensuremath{\pi}}\right)^{3}}e^{i\mathbf{k}\cdot\mathbf{r}} + \nonumber \\
& + \chi_{2D}\int\frac{dk_{z}}{2\pi}e^{ik_{z}z}\int\frac{d\mathbf{k}_{\parallel}}{\left(2\text{\ensuremath{\pi}}\right)^{2}}e^{i\mathbf{k_{\parallel}}\cdot\mathbf{r_{\parallel}}}k_{\parallel}^{2}V\left(\mathbf{k_{\parallel}},0\right).
\end{align}
Therefore
\begin{equation}
-\left(k_{\parallel}^{2}+k_{z}^{2}\right)V\left(\mathbf{k}\right)=-\frac{e}{\epsilon_{0}}+k_{\parallel}^{2}\chi_{2D}V\left(\mathbf{k}_{\parallel},z=0\right).
\end{equation}
Solving for $V\left(\mathbf{k}\right)$ we obtain
\begin{equation}
V\left(\mathbf{k}\right)=\frac{e}{\epsilon_{0}}\frac{1}{k_{\parallel}^{2}+k_{z}^{2}}-\frac{k_{\parallel}^{2}}{k_{\parallel}^{2}+k_{z}^{2}}\chi_{2D}V\left(\mathbf{k}_{\parallel},z=0\right).
\end{equation}
Fourier transforming the previous equation in the $k_{z}$ coordinate
(and taking $z=0$) we obtain
\begin{equation}
V\left(\mathbf{k}_{\parallel},z=0\right)=\frac{e}{2\pi\epsilon_{0}}\frac{\pi}{k_{\parallel}}-\frac{\pi}{2\pi}k_{\parallel}\chi_{2D}V\left(\mathbf{k}_{\parallel},z=0\right),
\end{equation}
where
\begin{equation}
V\left(\mathbf{k}_{\parallel},z=0\right)=\int_{-\infty}^{\infty}\frac{dk_{z}}{2\pi}V\left(\mathbf{k}_{\parallel},k_{z}\right).
\end{equation}
Solving for $V\left(\mathbf{k}_{\parallel},z=0\right)$ we obtain
\begin{align}
V\left(\mathbf{k}_{\parallel},z=0\right) & =\frac{e}{2\epsilon_{0}}\frac{1}{k_{\parallel}}\frac{1}{1+\chi_{2D}k_{\parallel}/2}\nonumber \\
& = \frac{e}{2\epsilon_{0}k_{\parallel}}\frac{1}{1+k_{\parallel}/\kappa_{\parallel}},
\end{align}
where $1/\kappa_{\parallel}=\chi_{2D}/2$. The Fourier transform
of the potential reads \cite{1979JETPL..29..658K}
\begin{align}
V\left(\mathbf{r}_{\parallel},z =0\right) & =\frac{e}{2\epsilon_{0}}\int\frac{d\theta dk_{\parallel}}{(2\pi)^{2}}\frac{e^{i\mathbf{k}_{\parallel}\cdot\mathbf{r}_{\parallel}}\kappa_{\parallel}}{\kappa_{\parallel}+k_{\parallel}} \nonumber \\
& =\frac{e\kappa_{\parallel}}{2\epsilon_{0}}\int_{0}^{\infty}\frac{dk_{\parallel}}{2\pi}\frac{J_{0}\left(k_{\parallel}r\right)}{\kappa_{\parallel}+k_{\parallel}} \nonumber\\
& = \frac{e\kappa_{\parallel}}{4\pi\epsilon_{0}}\frac{\pi}{2}\left[H_{0}\left(r\kappa_{\parallel}\right)-Y_{0}\left(r\kappa_{\parallel}\right)\right] \nonumber \\
& = \frac{e}{4\pi\epsilon_{0}r_{0}}\frac{\pi}{2}\left[H_{0}\left(\frac{r}{r_{0}}\right)-Y_{0}\left(\frac{r}{r_{0}}\right)\right],\label{eq:Keldysh_potential}
\end{align}
where $r_{0}=1/\kappa_\parallel$, and $H_{0}\left(x\right)$ and $Y_{0}\left(x\right)$ are the Struve function and the Bessel function of the second kind, respectively.

\begin{figure}
\includegraphics[scale=0.7]{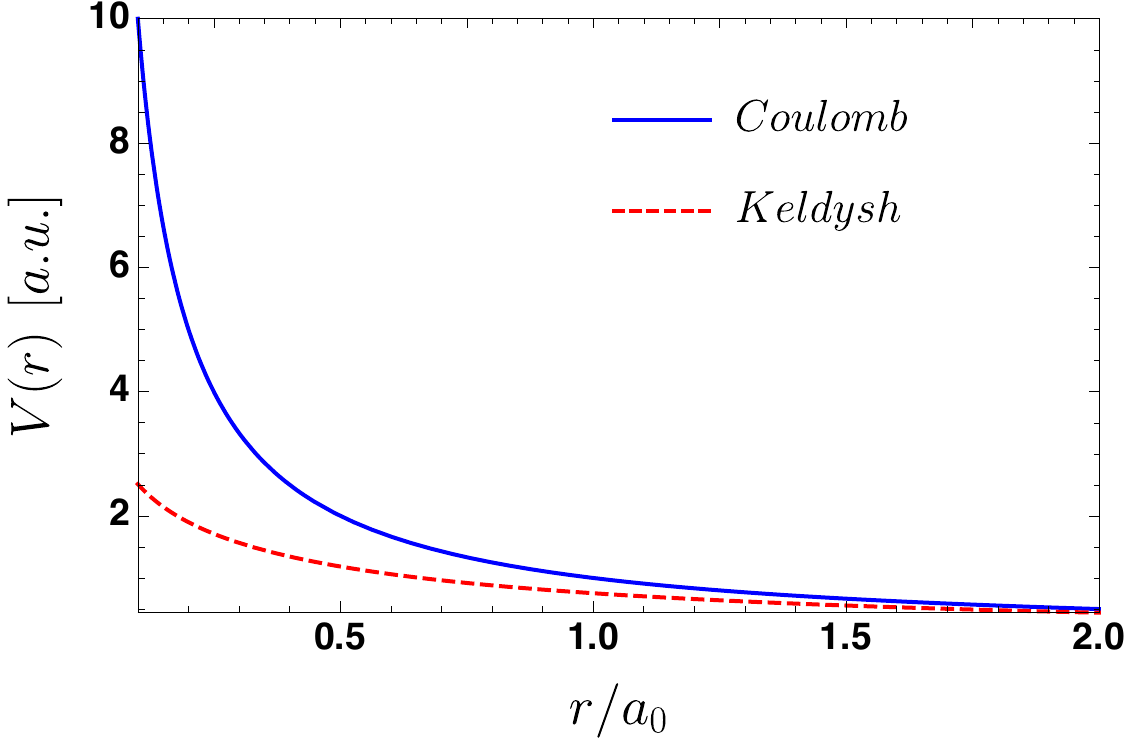}
\caption{Comparison of the Rytova-Keldysh and Coulomb potentials in arbitrary units.}
\end{figure}

\section{Wannier--Mott variational eigenvalue equation\label{sec:deriv_ham}}
In this appendix, we will quickly outline the necessary manipulations to obtain Eq. \ref{eq:variational_equation} from Eq. \ref{WM_ham}. Starting with $\left\{ \bra{\psi_{\mathrm{ex}}}\hat{H}\ket{\psi_{\mathrm{ex}}}\right\} _{\mathrm{th}}$, one substitutes in it the definition of Eq. \ref{eq:wannier}. Expanding the resulting equation, one immediately obtains the term proportional to the ground state energy, $E_{i} \sum_{k}\left|\phi\left(k\right)\right|^{2}$ when acting with the Hamiltonian on the ground state. Inspecting the remaining terms individually while utilizing the anti-commutation relations of the creation/destruction operators for each band, the terms proportional to $E_{c, \mathbf{k}}$ and $E_{v, \mathbf{k}}$ simplify almost immediately. 

Regarding the interaction terms, we start with 
\begin{equation}
V_{\mathbf{q}} c_{\mathbf{k}+\mathbf{q}}^{\dagger} b_{\mathbf{k}^{\prime}-\mathbf{q}}^{\dagger} b_{\mathbf{k}^{\prime}} c_{\mathbf{k}}.
\end{equation}
Contracting indices and, again, making use of the anti-commutation relations, this interaction term simplifies to 
\begin{align}
-\sum_{\mathbf{k}^{\prime \prime}} \phi^{*}\left(\mathbf{k}^{\prime \prime}\right) \frac{\sqrt{1-f_{c}\left(\mathbf{k}^{\prime \prime}\right)}}{A} \sum_{\mathbf{q} \neq 0} V_{\mathbf{q}} \phi\left(\mathbf{k}^{\prime \prime}-\mathbf{q}\right) \times & \nonumber \\ 
\times \sqrt{1-f_{c}\left(\mathbf{k}^{\prime \prime}-\mathbf{q}\right)}. &
\end{align}
The exchange terms for the conduction band, 
\begin{equation}
\frac{1}{2} V_{\mathbf{q}} c_{\mathbf{k}+\mathbf{q}}^{\dagger} c_{\mathbf{k}^{\prime}-\mathbf{q}}^{\dagger} c_{\mathbf{k}^{\prime}} c_{\mathbf{k}}, 
\end{equation}
result in 
\begin{align}
\frac{1}{2} \sum_{\mathbf{k}^{\prime}} \phi^{*}\left(\mathbf{k}^{\prime}\right) \frac{\sqrt{1-f_{c}\left(\mathbf{k}^{\prime}\right)}}{A} \sum_{\mathbf{q} \neq 0} V_{\mathbf{q}} \phi\left(\mathbf{k}^{\prime}-\mathbf{q}\right)\times & \nonumber \\
\times \sqrt{1-f_{c}\left(\mathbf{k}^{\prime}-\mathbf{q}\right)}, &
\end{align}
while those for the valence band, 
\begin{equation}
\frac{1}{2} V_{\mathbf{q}} b_{\mathbf{k}+\mathbf{q}}^{\dagger} b_{\mathbf{k}^{\prime}-\mathbf{q}}^{\dagger} b_{\mathbf{k}^{\prime}} b_{\mathbf{k}}, 
\end{equation}
result in
\begin{equation}
-\frac{1}{2} \sum_{\mathbf{k}} \sum_{\mathbf{q} \neq 0} V_{\mathbf{q}}\left|\phi\left(\mathbf{k}\right)\right|^{2}.
\end{equation}

Including these exchange terms in the definition of $\bar{E}_{c/v}\left(\mathbf{k}\right)$ (as seen in Eq. \ref{eq:variational_equation}), the eigenvalue equation becomes
\begin{align}
\sum_{\mathbf{k}}\phi^{*}\left(\mathbf{k}\right)\left\{\left[\bar{E}_{c, \mathbf{k}}-\bar{E}_{v, \mathbf{k}}\right] \phi\left(\mathbf{k}\right) + E_{i} \phi\left(\mathbf{k}\right) - \right. & \nonumber \\
\left. -\frac{\sqrt{1-f_{c}\left(\mathbf{k}\right)}}{A} \sum_{\mathbf{q}} V\left(\mathbf{q}\right) \phi\left(\mathbf{k}-\mathbf{q}\right) \sqrt{1-f_{c}\left(\mathbf{k}-\mathbf{q}\right)} \right\} = & \nonumber \\
=\sum_{\mathbf{k}} \phi^{*}\left(\mathbf{k}\right) E_{\mathrm{ex}} \phi\left(\mathbf{k}\right). &
\end{align}
Dropping the non-zero common factor $\phi^{*}\left(\mathbf{k}\right)$ while analyzing this sum term-by-term, we obtain the eigenvalue equation of Eq. \ref{eq:variational_equation} \begin{align}
\left[E_{\mathrm{ex}}-E_{i}\right]\phi\left(\mathbf{k}\right) =\left[\bar{E}_{c}\left(\mathbf{k}\right)-\bar{E}_{v}\left(\mathbf{k}\right)\right]\phi\left(\mathbf{k}\right)- & \nonumber\\
-\frac{\sqrt{1-f_{c}\left(\mathbf{k}\right)}}{A}\sum_{\mathbf{q}}V\left(\mathbf{q}\right)\phi\left(\mathbf{k}-\mathbf{q}\right)\sqrt{1-f_{c}\left(\mathbf{k}-\mathbf{q}\right)} &.
\end{align}

% BibTeX users please use
\bibliographystyle{unsrt}
%\bibliography{./PhD}

\begin{thebibliography}{100}

\bibitem{Mounet_2018}
Nicolas Mounet, Marco Gibertini, Philippe Schwaller, Davide Campi, Andrius
  Merkys, Antimo Marrazzo, Thibault Sohier, Ivano~Eligio Castelli, Andrea
  Cepellotti, Giovanni Pizzi, and Nicola Marzari.
\newblock Two-dimensional materials from high-throughput computational
  exfoliation of experimentally known compounds.
\newblock {\em Nat. Nano.}, 13(3):246--252, feb 2018.

\bibitem{1963}
R.~F. Frindt and A.~D. Yoffe.
\newblock Phys. properties of layer structures : optical properties and
  photoconductivity of thin crystals of molybdenum disulphide.
\newblock {\em Proc. R. Soc. Lond. A}, 273(1352):69--83, apr 1963.

\bibitem{Fortin_1982}
E.~Fortin and W.M. Sears.
\newblock Photovoltaic effect and optical absorption in {MoS}2.
\newblock {\em J. of Phys. and Chem. of Solids}, 43(9):881--884, jan 1982.

\bibitem{mueller_exciton_2018}
Thomas Mueller and Ermin Malic.
\newblock Exciton physics and device application of two-dimensional transition
  metal dichalcogenide semiconductors.
\newblock {\em npj {2D} Mat. and Applications}, 2(1):29, December 2018.

\bibitem{RevModPhys.90.021001}
Gang Wang, Alexey Chernikov, Mikhail~M. Glazov, Tony~F. Heinz, Xavier Marie,
  Thierry Amand, and Bernhard Urbaszek.
\newblock Colloquium: Excitons in atomically thin transition metal
  dichalcogenides.
\newblock {\em Rev. Mod. Phys.}, 90:021001, Apr 2018.

\bibitem{nwu078}
Hongyi Yu, Xiaodong Cui, Xiaodong Xu, and Wang Yao.
\newblock Valley excitons in two-dimensional semiconductors.
\newblock {\em National Science Review}, 2(1):57--70, 2015.

\bibitem{Knox_1983}
R.~S. Knox.
\newblock Introduction to exciton phys.
\newblock In {\em Collective Excitations in Solids}, pages 183--245. Springer
  {US}, 1983.

\bibitem{rytova1967}
N.S. Rytova, Alexey Chernikov, and Mikhail Glazov.
\newblock Screened potential of a point charge in a thin film.
\newblock {\em Moscow University Phys. Bulletin}, 3:30, 01 1967.

\bibitem{1979JETPL..29..658K}
L.~V. {Keldysh}.
\newblock Coulomb interaction in thin semiconductor and semimetal films.
\newblock {\em Sov. J. Exp. Theor. Phys. Lett.}, 29:658, June 1979.

\bibitem{Geim_2013}
A.~K. Geim and I.~V. Grigorieva.
\newblock Van der {Waals} heterostructures.
\newblock {\em Nature}, 499(7459):419--425, jul 2013.

\bibitem{Dong_2019}
Xi-Ying Dong, Run-Ze Li, Jia-Pei Deng, and Zi-Wu Wang.
\newblock Interlayer exciton-polaron effect in transition metal dichalcogenides
  van der {Waals} heterostructures.
\newblock {\em J. of Phys. and Chem. of Solids}, 134:1--4, nov 2019.

\bibitem{Thygesen_2017}
Kristian~Sommer Thygesen.
\newblock Calculating excitons, plasmons, and quasiparticles in {2D} materials
  and van der {Waals} heterostructures.
\newblock {\em {2D} Mater.}, 4(2):022004, jun 2017.

\bibitem{brunetti_optical_2018}
Matthew~N Brunetti, Oleg~L Berman, and Roman~Ya Kezerashvili.
\newblock Optical absorption by indirect excitons in a transition metal
  dichalcogenide/hexagonal boron nitride heterostructure.
\newblock {\em J. of Phys.: Cond. Matt.}, 30(22):225001, June 2018.

\bibitem{van_tuan_coulomb_2018}
Dinh Van~Tuan, Min Yang, and Hanan Dery.
\newblock The {Coulomb} interaction in monolayer transition-metal
  dichalcogenides.
\newblock {\em Phys. Rev. B}, 98(12):125308, September 2018.
\newblock arXiv: 1801.00477.

\bibitem{zhang_two-dimensional_2019}
J-Z Zhang and J-Z Ma.
\newblock Two-dimensional excitons in monolayer transition metal
  dichalcogenides from radial equation and variational calculations.
\newblock {\em J. of Phys.: Cond. Matt.}, 31(10):105702, March 2019.

\bibitem{Scharf_2019}
Benedikt Scharf, Dinh~Van Tuan, Igor {\v{Z}}uti{\'{c}}, and Hanan Dery.
\newblock Dynamical screening in monolayer transition-metal dichalcogenides and
  its manifestations in the exciton spectrum.
\newblock {\em J. Phys.: Cond. Matt.}, 31(20):203001, mar 2019.

\bibitem{kamban_interlayer_2020}
Høgni~C. Kamban and Thomas~G. Pedersen.
\newblock Interlayer excitons in van der {Waals} heterostructures: {Binding}
  energy, {Stark} shift, and field-induced dissociation.
\newblock {\em Scientific Reports}, 10(1):5537, December 2020.

\bibitem{Cavalcante_2018}
L.~S.~R. Cavalcante, A.~Chaves, B.~Van Duppen, F.~M. Peeters, and D.~R.
  Reichman.
\newblock Electrostatics of electron-hole interactions in van der {Waals}
  heterostructures.
\newblock {\em Phys. Rev. B}, 97(12), mar 2018.

\bibitem{novoselov_electric_2004}
K.~S. Novoselov, A.~K. Geim, S.~V. Morozov, D.~Jiang, Y.~Zhang, S.~V. Dubonos,
  I.~V. Grigorieva, and A.~A. Firsov.
\newblock Electric {Field} {Effect} in {Atomically} {Thin} {Carbon} {Films}.
\newblock {\em Science}, 306(5696):666--669, October 2004.

\bibitem{novoselov_two-dimensional_2005}
K.~S. Novoselov, A.~K. Geim, S.~V. Morozov, D.~Jiang, M.~I. Katsnelson, I.~V.
  Grigorieva, S.~V. Dubonos, and A.~A. Firsov.
\newblock Two-dimensional gas of massless {Dirac} fermions in graphene.
\newblock {\em Nature}, 438(7065):197--200, November 2005.

\bibitem{Mak2010}
Kin~Fai Mak, Changgu Lee, James Hone, Jie Shan, and Tony~F. Heinz.
\newblock Atomically thin {MoS2}: A new direct-gap semiconductor.
\newblock {\em Phys. Rev. Lett.}, 105(13):136805, September 2010.

\bibitem{wang_2010}
Andrea Splendiani, Liang Sun, Yuanbo Zhang, Tianshu Li, Jonghwan Kim, Chi-Yung
  Chim, Giulia Galli, and Feng Wang.
\newblock Emerging photoluminescence in monolayer {MoS2}.
\newblock {\em Nano Lett.}, 10(4):1271--1275, 2010.
\newblock PMID: 20229981.

\bibitem{Wurstbauer_2017}
Ursula Wurstbauer, Bastian Miller, Eric Parzinger, and Alexander~W Holleitner.
\newblock Light{\textendash}matter interaction in transition metal
  dichalcogenides and their heterostructures.
\newblock {\em J. Phys. D: Appl. Phys.}, 50(17):173001, mar 2017.

\bibitem{chernikov_2014}
Alexey Chernikov, Timothy~C. Berkelbach, Heather~M. Hill, Albert Rigosi, Yilei
  Li, Ozgur~Burak Aslan, David~R. Reichman, Mark~S. Hybertsen, and Tony~F.
  Heinz.
\newblock Exciton binding energy and nonhydrogenic rydberg series in monolayer
  {WS2}.
\newblock {\em Phys. Rev. Lett.}, 113:076802, Aug 2014.

\bibitem{Mak_2012}
Kin~Fai Mak, Keliang He, Changgu Lee, Gwan~Hyoung Lee, James Hone, Tony~F.
  Heinz, and Jie Shan.
\newblock Tightly bound trions in monolayer {MoS}2.
\newblock {\em Nat. Mat.}, 12(3):207--211, dec 2012.

\bibitem{He_2014}
Keliang He, Nardeep Kumar, Liang Zhao, Zefang Wang, Kin~Fai Mak, Hui Zhao, and
  Jie Shan.
\newblock Tightly bound excitons in monolayer {WSe}2.
\newblock {\em Phys. Rev. Lett.}, 113(2), jul 2014.

\bibitem{Xiao_2012}
Di~Xiao, Gui-Bin Liu, Wanxiang Feng, Xiaodong Xu, and Wang Yao.
\newblock Coupled spin and valley phys. in monolayers of {MoS}2 and other
  group-{VI} dichalcogenides.
\newblock {\em Phys. Rev. Lett.}, 108(19), may 2012.

\bibitem{Geim2013}
A.~K. Geim and I.~V. Grigorieva.
\newblock Van der {Waals} heterostructures.
\newblock {\em Nature}, 499(7459):419--425, jul 2013.

\bibitem{kis_2011}
Zongyou Yin, Hai Li, Hong Li, Lin Jiang, Yumeng Shi, Yinghui Sun, Gang Lu, Qing
  Zhang, Xiaodong Chen, and Hua Zhang.
\newblock Single-layer {MoS2} phototransistors.
\newblock {\em ACS Nano}, 6(1):74--80, 2012.
\newblock PMID: 22165908.

\bibitem{Yin_2011}
Zongyou Yin, Hai Li, Hong Li, Lin Jiang, Yumeng Shi, Yinghui Sun, Gang Lu, Qing
  Zhang, Xiaodong Chen, and Hua Zhang.
\newblock Single-layer {MoS}2 phototransistors.
\newblock {\em ACS Nano}, 6(1):74--80, dec 2011.

\bibitem{Pospischil_2014}
Andreas Pospischil, Marco~M. Furchi, and Thomas Mueller.
\newblock Solar-energy conversion and light emission in an atomic monolayer
  p{\textendash}n diode.
\newblock {\em Nat. Nano.}, 9(4):257--261, mar 2014.

\bibitem{Furchi2014}
Marco~M. Furchi, Andreas Pospischil, Florian Libisch, Joachim Burgd{\"o}rfer,
  and Thomas Mueller.
\newblock Photovoltaic effect in an electrically tunable van der {Waals}
  heterojunction.
\newblock {\em Nano Lett.}, 14(8):4785--4791, July 2014.

\bibitem{Amani_2015}
M.~Amani, D.-H. Lien, D.~Kiriya, J.~Xiao, A.~Azcatl, J.~Noh, S.~R. Madhvapathy,
  R.~Addou, S.~KC, M.~Dubey, K.~Cho, R.~M. Wallace, S.-C. Lee, J.-H. He, J.~W.
  Ager, X.~Zhang, E.~Yablonovitch, and A.~Javey.
\newblock Near-unity photoluminescence quantum yield in {MoS}2.
\newblock {\em Science}, 350(6264):1065--1068, nov 2015.

\bibitem{Zhang_2017}
Xiao-Xiao Zhang, Ting Cao, Zhengguang Lu, Yu-Chuan Lin, Fan Zhang, Ying Wang,
  Zhiqiang Li, James~C. Hone, Joshua~A. Robinson, Dmitry Smirnov, Steven~G.
  Louie, and Tony~F. Heinz.
\newblock Magnetic brightening and control of dark excitons in monolayer
  {WSe}2.
\newblock {\em Nat. Nano.}, 12(9):883--888, jun 2017.

\bibitem{Molas_2017}
M~R Molas, C~Faugeras, A~O Slobodeniuk, K~Nogajewski, M~Bartos, D~M Basko, and
  M~Potemski.
\newblock Brightening of dark excitons in monolayers of semiconducting
  transition metal dichalcogenides.
\newblock {\em {2D} Mat.}, 4(2):021003, jan 2017.

\bibitem{scharf_excitonic_2016}
Benedikt Scharf, Tobias Frank, Martin Gmitra, Jaroslav Fabian, Igor Žutić,
  and Vasili Perebeinos.
\newblock Excitonic {Stark} effect in {MoS}2 monolayers.
\newblock {\em Phys. Rev. B}, 94(24):245434, December 2016.

\bibitem{haastrup_stark_2016}
Sten Haastrup, Simone Latini, Kirill Bolotin, and Kristian~S. Thygesen.
\newblock Stark shift and electric-field-induced dissociation of excitons in
  monolayer {MoS}2 and h{BN}/{MoS}2 heterostructures.
\newblock {\em Phys. Rev. B}, 94(4):041401, July 2016.

\bibitem{massicotte_dissociation_2018}
Mathieu Massicotte, Fabien Vialla, Peter Schmidt, Mark~B. Lundeberg, Simone
  Latini, Sten Haastrup, Mark Danovich, Diana Davydovskaya, Kenji Watanabe,
  Takashi Taniguchi, Vladimir~I. Fal’ko, Kristian~S. Thygesen, Thomas~G.
  Pedersen, and Frank H.~L. Koppens.
\newblock Dissociation of two-dimensional excitons in monolayer {WSe2}.
\newblock {\em Nat. Comm.}, 9(1):1633, December 2018.

\bibitem{Rigosi_2015}
Albert~F. Rigosi, Heather~M. Hill, Yilei Li, Alexey Chernikov, and Tony~F.
  Heinz.
\newblock Probing interlayer interactions in transition metal dichalcogenide
  heterostructures by optical spectroscopy: {MoS}2/{WS}2 and {MoSe}2/{WSe}2.
\newblock {\em Nano Lett.}, 15(8):5033--5038, jul 2015.

\bibitem{Berkelbach_2018}
Timothy~C. Berkelbach and David~R. Reichman.
\newblock Optical and excitonic properties of atomically thin transition-metal
  dichalcogenides.
\newblock {\em Annu. Rev. Cond. Matt. Phys.}, 9(1):379--396, mar 2018.

\bibitem{aquino_accurate_2005}
N.~Aquino, G.~Campoy, and A.~Flores-Riveros.
\newblock Accurate energy eigenvalues and eigenfunctions for the
  two-dimensional confined hydrogen atom.
\newblock {\em Int. J. Quant. Chem.}, 103(3):267--277, 2005.

\bibitem{grasselli_variational_2017}
Federico Grasselli.
\newblock Variational approach to the soft-{Coulomb} potential in
  low-dimensional quantum systems.
\newblock {\em Am. J. Phys.}, 85(11):834--839, November 2017.

\bibitem{inarrea_effects_2019}
Manuel Iñarrea, Víctor Lanchares, Jesús~F. Palacián, Ana~I. Pascual,
  J.~Pablo Salas, and Patricia Yanguas.
\newblock Effects of a soft-core coulomb potential on the dynamics of a
  hydrogen atom near a metal surface.
\newblock {\em Comm. in Nonlinear Science and Numerical Simulation},
  68:94--105, March 2019.

\bibitem{Shahnazaryan_2017}
V.~Shahnazaryan, I.~Iorsh, I.~A. Shelykh, and O.~Kyriienko.
\newblock Exciton-exciton interaction in transition-metal dichalcogenide
  monolayers.
\newblock {\em Phys. Rev. B}, 96(11), sep 2017.

\bibitem{Aas_2018}
Shahnaz Aas and Ceyhun Bulutay.
\newblock Strain dependence of photoluminescence and circular dichroism in
  transition metal dichalcogenides: a k$\cdotp$p analysis.
\newblock {\em Opt. Express}, 26(22):28672, oct 2018.

\bibitem{Jin_2019}
Chenhao Jin, Emma~C. Regan, Aiming Yan, M.~Iqbal~Bakti Utama, Danqing Wang,
  Sihan Zhao, Ying Qin, Sijie Yang, Zhiren Zheng, Shenyang Shi, Kenji Watanabe,
  Takashi Taniguchi, Sefaattin Tongay, Alex Zettl, and Feng Wang.
\newblock Observation of moir{\'{e}} excitons in {WSe}2/{WS}2 heterostructure
  superlattices.
\newblock {\em Nature}, 567(7746):76--80, feb 2019.

\bibitem{Castellanos_Gomez_2015}
Andres Castellanos-Gomez.
\newblock Black phosphorus: Narrow gap, wide applications.
\newblock {\em J. Phys. Chem. Lett.}, 6(21):4280--4291, oct 2015.

\bibitem{Li_2015}
M.-Y. Li, Y.~Shi, C.-C. Cheng, L.-S. Lu, Y.-C. Lin, H.-L. Tang, M.-L. Tsai,
  C.-W. Chu, K.-H. Wei, J.-H. He, W.-H. Chang, K.~Suenaga, and L.-J. Li.
\newblock Epitaxial growth of a monolayer {WSe}2-{MoS}2 lateral p-n junction
  with an atomically sharp interface.
\newblock {\em Science}, 349(6247):524--528, jul 2015.

\bibitem{Pawbake_2016}
Amit~S. Pawbake, Mahendra~S. Pawar, Sandesh~R. Jadkar, and Dattatray~J. Late.
\newblock Large area chemical vapor deposition of monolayer transition metal
  dichalcogenides and their temperature dependent raman spectroscopy studies.
\newblock {\em Nanoscale}, 8(5):3008--3018, 2016.

\bibitem{Choi_2017}
Wonbong Choi, Nitin Choudhary, Gang~Hee Han, Juhong Park, Deji Akinwande, and
  Young~Hee Lee.
\newblock Recent development of two-dimensional transition metal
  dichalcogenides and their applications.
\newblock {\em Materials Today}, 20(3):116--130, apr 2017.

\bibitem{Ratnikov_2020}
Pavel~V. Ratnikov.
\newblock Excitons in planar quantum wells based on transition metal
  dichalcogenides.
\newblock {\em Phys. Rev. B}, 102(8), aug 2020.

\bibitem{chaves_excitonic_2017}
A~J Chaves, R~M Ribeiro, T~Frederico, and N~M~R Peres.
\newblock Excitonic effects in the optical properties of {2D} materials: an
  equation of motion approach.
\newblock {\em {2D} Mat.}, 4(2):025086, April 2017.

\bibitem{have_excitonic_2019}
J.~Have, N.~M.~R. Peres, and T.~G. Pedersen.
\newblock Excitonic magneto-optics in monolayer transition metal
  dichalcogenides: {From} nanoribbons to two-dimensional response.
\newblock {\em Phys. Rev. B}, 100(4):045411, July 2019.

\bibitem{henriques_excitonic_2020}
J.~C.~G. Henriques, G.~Catarina, A.~T. Costa, J.~Fernández-Rossier, and
  N.~M.~R. Peres.
\newblock Excitonic magneto-optical {Kerr} effect in two-dimensional transition
  metal dichalcogenides induced by spin proximity.
\newblock {\em Phys. Rev. B}, 101(4):045408, January 2020.

\bibitem{henriques_optical_2020}
J~C~G Henriques, G~B Ventura, C~D~M Fernandes, and N~M~R Peres.
\newblock Optical absorption of single-layer hexagonal boron nitride in the
  ultraviolet.
\newblock {\em J. of Phys.: Cond. Matt.}, 32(2):025304, January 2020.

\bibitem{durnev_excitons_nodate}
M~V Durnev and M~M Glazov.
\newblock Excitons and trions in two-dimensional semiconductors based on
  transition metal dichalcogenides.
\newblock {\em Physics-Uspekhi}, 61(9):825--845, sep 2018.

\bibitem{van_der_donck_excitons_2018}
M.~Van~der Donck, M.~Zarenia, and F.~M. Peeters.
\newblock Excitons, trions, and biexcitons in transition-metal dichalcogenides:
  {Magnetic}-field dependence.
\newblock {\em Phys. Rev. B}, 97(19):195408, May 2018.

\bibitem{Castellanos_Gomez_2014}
Andres Castellanos-Gomez, Leonardo Vicarelli, Elsa Prada, Joshua~O Island, K~L
  Narasimha-Acharya, Sofya~I Blanter, Dirk~J Groenendijk, Michele Buscema,
  Gary~A Steele, J~V Alvarez, Henny~W Zandbergen, J~J Palacios, and Herre S~J
  van~der Zant.
\newblock Isolation and characterization of few-layer black phosphorus.
\newblock {\em {2D} Mater.}, 1(2):025001, jun 2014.

\bibitem{Zhang_2019}
J-Z Zhang and J-Z Ma.
\newblock Two-dimensional excitons in monolayer transition metal
  dichalcogenides from radial equation and variational calculations.
\newblock {\em J. Phys.: Cond. Matt.}, 31(10):105702, jan 2019.

\bibitem{Stark_1914}
J.~Stark.
\newblock Beobachtungen über den effekt des elektrischen feldes auf
  spektrallinien. i. quereffekt.
\newblock {\em Ann. Phys.}, 348(7):965--982, 1914.

\bibitem{Simon_1979}
Barry Simon.
\newblock The definition of molecular resonance curves by the method of
  exterior complex scaling.
\newblock {\em Phys. Lett. A}, 71(2-3):211--214, apr 1979.

\bibitem{McCurdy_1991}
C.~William McCurdy, Carrie~K. Stroud, and Matthew~K. Wisinski.
\newblock Solving the time-dependent schrödinger equation using
  complex-coordinate contours.
\newblock {\em Phys. Rev. A}, 43(11):5980--5990, jun 1991.

\bibitem{Hong_2014}
Xiaoping Hong, Jonghwan Kim, Su-Fei Shi, Yu~Zhang, Chenhao Jin, Yinghui Sun,
  Sefaattin Tongay, Junqiao Wu, Yanfeng Zhang, and Feng Wang.
\newblock Ultrafast charge transfer in atomically thin {MoS}2/{WS}2
  heterostructures.
\newblock {\em Nat. Nano.}, 9(9):682--686, aug 2014.

\bibitem{Meckbach_2018}
L.~Meckbach, T.~Stroucken, and S.~W. Koch.
\newblock Influence of the effective layer thickness on the ground-state and
  excitonic properties of transition-metal dichalcogenide systems.
\newblock {\em Phys. Rev. B}, 97(3), jan 2018.

\bibitem{Kumar_2018}
Rajeev Kumar, Ivan Verzhbitskiy, Francesco Giustiniano, Themistoklis P~H
  Sidiropoulos, Rupert~F Oulton, and Goki Eda.
\newblock Interlayer screening effects in a van der {Waals} heterobilayer.
\newblock {\em {2D} Mater.}, 5(4):041003, aug 2018.

\bibitem{Vialla_2019}
Fabien Vialla, Mark Danovich, David~A Ruiz-Tijerina, Mathieu Massicotte, Peter
  Schmidt, Takashi Taniguchi, Kenji Watanabe, Ryan~J Hunt, Marcin Szyniszewski,
  Neil~D Drummond, Thomas~G Pedersen, Vladimir~I Fal'ko, and Frank H~L Koppens.
\newblock Tuning of impurity-bound interlayer complexes in a van der {Waals}
  heterobilayer.
\newblock {\em {2D} Mater.}, 6(3):035032, may 2019.

\bibitem{Malic-2019}
Simon Ovesen, Samuel Brem, Christopher Linder\"{a}lv, Mikael Kuisma, Tobias
  Korn, Paul Erhart, Malte Selig, and Ermin Malic.
\newblock Interlayer exciton dynamics in van der {Waals} heterostructures.
\newblock {\em Communications Physics}, 2(1), February 2019.

\bibitem{Miller_2017}
Bastian Miller, Alexander Steinhoff, Borja Pano, Julian Klein, Frank Jahnke,
  Alexander Holleitner, and Ursula Wurstbauer.
\newblock Long-lived direct and indirect interlayer excitons in van der {Waals}
  heterostructures.
\newblock {\em Nano Lett.}, 17(9):5229--5237, aug 2017.

\bibitem{Kamban_2019}
H{\o}gni~C. Kamban and Thomas~G. Pedersen.
\newblock Field-induced dissociation of two-dimensional excitons in transition
  metal dichalcogenides.
\newblock {\em Phys. Rev. B}, 100(4), jul 2019.

\bibitem{Wang_2016}
Haining Wang, Changjian Zhang, Weimin Chan, Christina Manolatou, Sandip Tiwari,
  and Farhan Rana.
\newblock Radiative lifetimes of excitons and trions in monolayers of the metal
  dichalcogenide {MoS}2.
\newblock {\em Phys. Rev. B}, 93(4), jan 2016.

\bibitem{Moody_2015}
Galan Moody, Chandriker~Kavir Dass, Kai Hao, Chang-Hsiao Chen, Lain-Jong Li,
  Akshay Singh, Kha Tran, Genevieve Clark, Xiaodong Xu, Gunnar Berghäuser,
  Ermin Malic, Andreas Knorr, and Xiaoqin Li.
\newblock Intrinsic homogeneous linewidth and broadening mechanisms of excitons
  in monolayer transition metal dichalcogenides.
\newblock {\em Nat. Comm.}, 6(1), sep 2015.

\bibitem{Han_2018}
B.~Han, C.~Robert, E.~Courtade, M.~Manca, S.~Shree, T.~Amand, P.~Renucci,
  T.~Taniguchi, K.~Watanabe, X.~Marie, L.{\hspace{0.167em}}E. Golub,
  M.{\hspace{0.167em}}M. Glazov, and B.~Urbaszek.
\newblock Exciton states in monolayer {MoSe}2 and {MoTe}2 probed by
  upconversion spectroscopy.
\newblock {\em Phys. Rev. X}, 8(3), sep 2018.

\bibitem{Courtade_2017}
E.~Courtade, M.~Semina, M.~Manca, M.~M. Glazov, C.~Robert, F.~Cadiz, G.~Wang,
  T.~Taniguchi, K.~Watanabe, M.~Pierre, W.~Escoffier, E.~L. Ivchenko,
  P.~Renucci, X.~Marie, T.~Amand, and B.~Urbaszek.
\newblock Charged excitons in monolayer {WSe}2: Experiment and theory.
\newblock {\em Phys. Rev. B}, 96(8), aug 2017.

\bibitem{zhang_absorption_2014}
Changjian Zhang, Haining Wang, Weimin Chan, Christina Manolatou, and Farhan
  Rana.
\newblock Absorption of light by excitons and trions in monolayers of metal
  dichalcogenide {Mo}{S}2: {Experiments} and theory.
\newblock {\em Phys. Rev. B}, 89(20):205436, May 2014.

\bibitem{van_der_donck_excitons_2017}
M.~Van~der Donck, M.~Zarenia, and F.~M. Peeters.
\newblock Excitons and trions in monolayer transition metal dichalcogenides:
  {A} comparative study between the multiband model and the quadratic
  single-band model.
\newblock {\em Phys. Rev. B}, 96(3):035131, July 2017.

\bibitem{van_der_donck_interlayer_2018}
M.~Van~der Donck and F.~M. Peeters.
\newblock Interlayer excitons in transition metal dichalcogenide
  heterostructures.
\newblock {\em Phys. Rev. B}, 98(11):115104, September 2018.

\bibitem{Ruppert-2014}
Claudia Ruppert, Ozgur~Burak Aslan, and Tony~F. Heinz.
\newblock Optical properties and band gap of single- and few-layer {MoTe2}
  crystals.
\newblock 14(11):6231, 2014.

\bibitem{Semina_2019}
M.~A. Semina.
\newblock Excitons and trions in bilayer van der {Waals} heterostructures.
\newblock {\em Phys. Sol. Stat.}, 61(11):2218--2223, nov 2019.

\bibitem{planelles_simple_2017}
Josep Planelles.
\newblock Simple correlated wave-function for excitons in {0D}, quasi-{1D} and
  quasi-{2D} quantum dots.
\newblock {\em Theor. Chem. Accounts}, 136(7):81, July 2017.

\bibitem{Lundt_2018}
N.~Lundt, E.~Cherotchenko, O.~Iff, X.~Fan, Y.~Shen, P.~Bigenwald, A.~V.
  Kavokin, S.~Höfling, and C.~Schneider.
\newblock The interplay between excitons and trions in a monolayer of {MoSe}2.
\newblock {\em Appl. Phys. Lett.}, 112(3):031107, jan 2018.

\bibitem{bhat_flexural_1987}
R.B. Bhat.
\newblock Flexural vibration of polygonal plates using characteristic
  orthogonal polynomials in two variables.
\newblock {\em J. of Sound and Vibration}, 114(1):65--71, January 1987.

\bibitem{liew_set_1991}
K.~M. Liew and K.~Y. Lam.
\newblock A {Set} of {Orthogonal} {Plate} {Functions} for {Flexural}
  {Vibration} of {Regular} {Polygonal} {Plates}.
\newblock {\em J. of Vibration and Acoustics}, 113(2):182--186, April 1991.

\bibitem{Quintela_2020}
M~F C~Martins Quintela and J~M B~Lopes dos Santos.
\newblock A polynomial approach to the spectrum of dirac{\textendash}weyl
  polygonal billiards.
\newblock {\em Journal of Physics: Condensed Matter}, 33(3):035901, oct 2020.

\bibitem{Brown_1987}
Jerry~W. Brown and Harold~N. Spector.
\newblock Exciton binding energy in a quantum-well wire.
\newblock {\em Phys. Rev. B}, 35(6):3009--3012, feb 1987.

\bibitem{Feng_1993}
Yuan ping Feng and Harold~N. Spector.
\newblock Exciton energies as a function of electric field: Confined quantum
  stark effect.
\newblock {\em Phys. Rev. B}, 48(3):1963--1966, jul 1993.

\bibitem{Akimoto_1972}
Okikazu Akimoto and Eiichi Hanamura.
\newblock Excitonic molecule. {I}. {Calculation} of the binding energy.
\newblock {\em J. Phys. Soc. Jpn.}, 33(6):1537--1544, dec 1972.

\bibitem{Loudon_1959}
Rodney Loudon.
\newblock One-dimensional hydrogen atom.
\newblock {\em Am. J. Phys.}, 27(9):649--655, dec 1959.

\bibitem{Henriques_2019}
J~C~G Henriques, G~B Ventura, C~D~M Fernandes, and N~M~R Peres.
\newblock Optical absorption of single-layer hexagonal boron nitride in the
  ultraviolet.
\newblock {\em J. Phys.: Cond. Matt.}, 32(2):025304, oct 2019.

\bibitem{PhysRevB.84.085406}
Pierluigi Cudazzo, Ilya~V. Tokatly, and Angel Rubio.
\newblock Dielectric screening in two-dimensional insulators: Implications for
  excitonic and impurity states in graphane.
\newblock {\em Phys. Rev. B}, 84:085406, Aug 2011.

\bibitem{Schmitt_Rink_1985}
S.~Schmitt-Rink and C.~Ell.
\newblock Excitons and electron-hole plasma in quasi-two-dimensional systems.
\newblock {\em J. of Luminescence}, 30(1-4):585--596, feb 1985.

\bibitem{kamban2020anisotropic}
Høgni~C. Kamban, Thomas~G. Pedersen, and Nuno M.~R. Peres.
\newblock Anisotropic stark shift, field-induced dissociation, and
  electroabsorption of excitons in phosphorene.
\newblock {\em Phys. Rev. B}, 102:115305, September 2020.

\bibitem{Wannier_1937}
Gregory~H. Wannier.
\newblock The structure of electronic excitation levels in insulating crystals.
\newblock {\em Phys. Rev.}, 52(3):191--197, aug 1937.

\bibitem{mott_basis_1949}
N.~F. Mott.
\newblock The {Basis} of the {Electron} {Theory} of {Metals}, with {Special}
  {Reference} to the {Transition} {Metals}.
\newblock {\em Proc. Phys. Soc.. Sec. A}, 62(7):416--422, July 1949.

\bibitem{Have_2019}
J.~Have, G.~Catarina, T.~G. Pedersen, and N.~M.~R. Peres.
\newblock Monolayer transition metal dichalcogenides in strong magnetic fields:
  Validating the wannier model using a microscopic calculation.
\newblock {\em Phys. Rev. B}, 99(3), jan 2019.

\bibitem{Chang_2019}
Yao-Wen Chang and David~R. Reichman.
\newblock Many-body theory of optical absorption in doped two-dimensional
  semiconductors.
\newblock {\em Phys. Rev. B}, 99(12), mar 2019.

\bibitem{Van_Tuan_2019}
Dinh~Van Tuan, Benedikt Scharf, Zefang Wang, Jie Shan, Kin~Fai Mak, Igor
  {\v{Z}}uti{\'{c}}, and Hanan Dery.
\newblock Probing many-body interactions in monolayer transition-metal
  dichalcogenides.
\newblock {\em Phys. Rev. B}, 99(8), feb 2019.

\bibitem{mahan_many-particle-physics}
Gerald~D. Mahan.
\newblock {\em Many-particle physics}.
\newblock Physics of solids and liquids. Plenum Press, 2nd ed edition, 1990.

\bibitem{Haug_2009}
Stephan W.~Koch Hartmut~Haug.
\newblock {\em Quantum Theory of Optical and Electronic Properties of
  Semiconductors}.
\newblock World Scientific Publishing Company, 4 edition, 2004.

\bibitem{Ouerdane_2011}
H.~Ouerdane.
\newblock Analytic model of effective screened coulomb interactions in a
  multilayer system.
\newblock {\em J. Appl. Phys.}, 110(7):074905, oct 2011.

\bibitem{Peres_2010}
N.~M.~R. Peres, R.~M. Ribeiro, and A.~H.~Castro Neto.
\newblock Excitonic effects in the optical conductivity of gated graphene.
\newblock {\em Phys. Rev. Lett.}, 105(5), jul 2010.

\bibitem{Glutsch_2004}
Stephan Glutsch.
\newblock {\em Excitons in Low-Dimensional Semiconductors}.
\newblock Springer Berlin Heidelberg, 2004.

\bibitem{Berkelbach_2013}
Timothy~C. Berkelbach, Mark~S. Hybertsen, and David~R. Reichman.
\newblock Theory of neutral and charged excitons in monolayer transition metal
  dichalcogenides.
\newblock {\em Phys. Rev. B}, 88(4), jul 2013.

\bibitem{Cudazzo_2011}
Pierluigi Cudazzo, Ilya~V. Tokatly, and Angel Rubio.
\newblock Dielectric screening in two-dimensional insulators: Implications for
  excitonic and impurity states in graphane.
\newblock {\em Phys. Rev. B}, 84(8), aug 2011.

\bibitem{Qiu_2013}
Diana~Y. Qiu, Felipe~H. da~Jornada, and Steven~G. Louie.
\newblock Optical spectrum of {MoS}2: Many-body effects and diversity of
  exciton states.
\newblock {\em Phys. Rev. Lett.}, 111(21), nov 2013.

\bibitem{avalos-ovando_lateral_2019}
O~Ávalos Ovando, D~Mastrogiuseppe, and S~E Ulloa.
\newblock Lateral heterostructures and one-dimensional interfaces in {2D}
  transition metal dichalcogenides.
\newblock {\em J. of Phys.: Cond. Matt.}, 31(21):213001, May 2019.

\bibitem{snoke2009solid}
D.~W. Snoke.
\newblock {\em Solid state physics : essential concepts}.
\newblock Addison-Wesley, San Francisco, 2009.

\bibitem{jackson1999classical}
John Jackson.
\newblock {\em Classical electrodynamics}.
\newblock Wiley, New York, 1999.

\bibitem{Berman-2017}
Oleg~L. Berman, Godfrey Gumbs, and Roman~Ya. Kezerashvili.
\newblock Bose-{Einstein} condensation and superfluidity of dipolar excitons in
  a phosphorene double layer.
\newblock {\em Phys. Rev. B}, 96:014505, Jul 2017.

\bibitem{Edery_2018}
Ariel Edery and Philippe Laporte.
\newblock First and second-order relativistic corrections to the two and
  higher-dimensional isotropic harmonic oscillator obeying the spinless
  salpeter equation.
\newblock {\em J. Phys. Comm.}, 2(2):025024, feb 2018.

\bibitem{cohen-tannoudji_quantum_2019}
Bernard Cohen-Tannoudji, Frank Laloe, and Diu Claude.
\newblock {\em {Quantum} {Mechanics}: volume 3.}
\newblock Wiley, 12 2019.

\bibitem{Horng_2018}
Jason Horng, Tineke Stroucken, Long Zhang, Eunice~Y. Paik, Hui Deng, and
  Stephan~W. Koch.
\newblock Observation of interlayer excitons in {MoSe}2 single crystals.
\newblock {\em Phys. Rev. B}, 97(24), jun 2018.

\bibitem{Rahaman2019}
Mahfujur Rahaman, Christian Wagner, Ashutosh Mukherjee, Adan Lopez-Rivera,
  Sibylle Gemming, and Dietrich R~T Zahn.
\newblock Probing interlayer excitons in a vertical van der waals p-n junction
  using a scanning probe microscopy technique.
\newblock {\em Journal of Physics: Condensed Matter}, 31(11):114001, January
  2019.

\bibitem{Kyl_np__2015}
Ilkka Kylänpää and Hannu-Pekka Komsa.
\newblock Binding energies of exciton complexes in transition metal
  dichalcogenide monolayers and effect of dielectric environment.
\newblock {\em Phys. Rev. B}, 92(20), nov 2015.

\bibitem{Lau_2019}
Ka~Wai Lau, Caterina Cocchi, and Claudia Draxl.
\newblock Electronic and optical excitations of two-dimensional {ZrS}2 and
  {HfS}2 and their heterostructure.
\newblock {\em Phys. Rev. Mat.}, 3(7), jul 2019.

\bibitem{0080168469}
G.F. Bassani, G.P. Parravicini, and R.A. Ballinger.
\newblock {\em Electronic States and Optical Transitions in Solids}.
\newblock International Series of Monographs on solid state physics. Pergamon
  Press, New York, 1975.

\end{thebibliography}
%

\end{document}